\documentclass{aa}
\usepackage{epsfig}
\usepackage{natbib}
\bibpunct{(}{)}{;}{\  a}{~}{,\ }
\begin{document}

   \title{
           ADAS analysis of the differential emission measure
           structure of the inner solar corona
         }
\subtitle{
	   II. A study of the `quiet Sun' inhomogeneities from 
           SOHO CDS-NIS spectra
         }

   \author{	A. C. Lanzafame \inst{1} 
	   	\and D. H. Brooks\inst{2}
	   	\and J. Lang\inst{3}
	  }
  \institute{
			Dipartimento di Fisica e Astronomia,
		       	Universit\`a di Catania,
		        Via S. Sofia 78,
		        I-95123 Catania, Italy\\
                        \email{alanzafame@ct.astro.it}
  \and
                	Kwasan Observatory,
                	Kyoto University,
                	Yamashina,	
               	 	Kyoto 607-8471,	
                	Japan
  \and
                        Rutherford Appleton Laboratory, 
                        Chilton, Didcot,
                        OX11 0QX, U.K.
             }

   \offprints{A. C. Lanzafame}

   \date{Received 4 October 2004 / Accepted 20 November 2004}

   \authorrunning{A. C. Lanzafame et al.}

   \titlerunning{DEM analysis of the CDS-NIS spectrum}

\abstract{We present a study of the differential emission measure
          (DEM) of a `quiet Sun' area observed in the extreme
          ultraviolet at normal incidence by the Coronal Diagnostic
          Spectrometer (CDS) on the SOHO spacecraft. The data used for
          this work were taken using the NISAT\_S observing
          sequence. This takes the full wavelength ranges from both
          the NIS channels (308-381\,\AA\ and 513-633\,\AA) with the 2
          arcsec by 240 arcsec slit, which is the narrowest slit
          available, yielding the best spectral resolution. In this
          work we contrast the DEM from subregions of 2$\times$80
          arcsec$^2$ with that obtained from the mean spectrum of the
          whole raster (20$\times$240 arcsec$^2$). We find that the
          DEM maintains essentially the same shape in the subregions,
          differing by a constant factor between 0.5 and 2 from the
          mean DEM, except in areas were the electron density is below
          $2 \times 10^7$ cm$^{-3}$ and downflow velocities of 50
          km\,s$^{-1}$ are found in the transition region. Such areas
          are likely to contain plasma departing from ionisation
          equilibrium, violating the basic assumptions underlying the
          DEM method.  The comparison between lines of Li-like and
          Be-like ions may provide further evidence of departure from
          ionisation equilibrium.  We find also that line intensities
          tend to be lower where velocities of the order of 30 km
          s$^{-1}$ or higher are measured in transition region
          lines. The DEM analysis is also exploited to improve the
          line identification performed by \citet{Brooks_etal:99} and
          to investigate possible elemental abundance variations from
          region to region. We find that the plasma has composition
          close to photospheric in all the subregions examined.
          \keywords{ Sun: atmosphere -- Sun: corona -- Sun: UV
          radiation -- Sun: abundances -- Atomic data -- Methods: data
          analysis -- Techniques: spectroscopic }}

\maketitle

\section{Introduction}

The discovery of highly dynamical phenomena in the `quiet Sun' (QS),
like blinkers \citep{Harrison:97,Harrison_etal:99} or explosive events
\citep{Innes_etal:97}, place further questions on the validity of
analysis methods which, like the DEM, rely on plasma equilibrium.
Indeed, many other small-scale transient phenomena have been reported
in the literature, including an attempt to obtain a global view of
such processes \citep{Harrison_etal:03} and for all such events care
should be taken to check the validity of the analysis method,
particularly whenever a DEM is invoked.  The controversy on the validity
of analysis methods based on emission measure is rooted, in fact, both
on the ill-posed nature of the mathematical problem and on the
assumption of ionisation equilibrium, which can be violated by the
evolution of the plasma on time-scales comparable to the
ionisation/recombination time-scales. In addition to these concerns,
the DEM can be rendered unreliable due to injudicious selection of
emission lines, for example, those affected by opacity, or those for
which the atomic coefficients are less accurate or incorrectly
modelled.

It is crucial, therefore, to understand {\it where} and {\it when}
emission measure type analyses can give reliable results. Answering
such questions has important consequences for establishing the
reliability of the physical quantities and the atmospheric morphology
extracted from observational data.

\citet{Lanzafame_etal:02}, hereafter Paper I, have studied the
validity and limitations of the DEM method in abundance determination,
spectral line identification, intensity predictions, and validation of
atomic cross-sections by analysing the SERTS-89 active region (AR)
spectrum \citep{Thomas_Neupert:94}. The integral inversion to infer
the DEM distribution from spectral line intensities was performed by
the {\it data adaptive smoothing approach}
\citep{Thompson:90,Thompson:91}, using an analysis procedure
integrated in ADAS (the Atomic Data and Analysis Structure). ADAS
allows a detailed atomic modelling which, notably, takes into account
the density dependence of both ionisation fractions and excitation
coefficients according to the {\it generalised collisional radiative
theory} \citep{McWhirter_Summers:84, Summers:94, Summers:01}. Such
data have not generally been included in DEM analyses in the past,
resulting in some confusion as to the interpretation of the real
reasons for discrepancies found in these analyses.

The study carried out in Paper I also showed that spurious multiple
peaks in the DEM distribution may derive from an inaccurate treatment
of the population densities of the excited levels and ionisation
fractions or from integral inversion techniques with arbitrary
smoothing. Complex DEM structures, such as those proposed for solar
and stellar coronae by several authors, have therefore been
questioned.

\begin{figure}[t]
\hbox to \hsize {\hss \epsfxsize=88mm \epsfbox {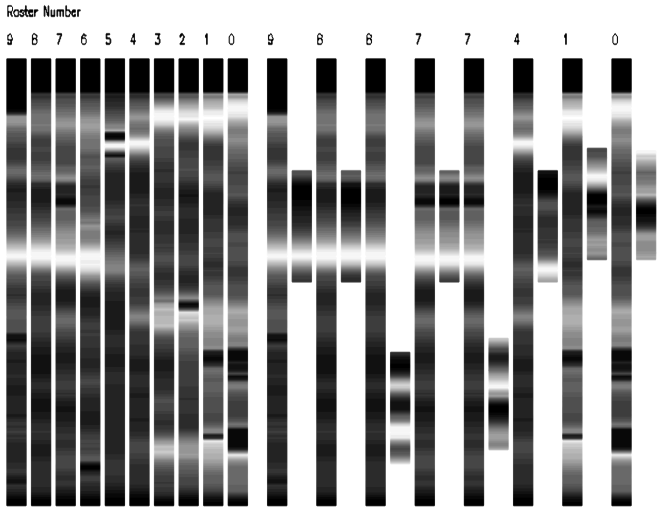}\hss}

   \caption{\ion{O}{v}\,(629.732\,\AA) image of the chosen dataset
   (s6011r00). It shows on the left the ten raster steps and on the
   right the positions of the 8 sements each 40 pixels long alongside
   all the pixels of the relevant raster step. These are denoted as
   follows: the segment from step 0 as 1(0); the segment from step 1
   as 2(1); the segment from step 4 as 3(4); the two segments from
   step 7 as 4(7L) for the lower one and 5(7U) for the upper one; the
   two segments from step 8 as 6(8L) for the lower one and 7(8U) for
   the upper one; the segment from step 9 as 8(9). }
\label{fig:OV_raster}
\end{figure}

Understanding {\it how} inhomogeneities combine to produce an observed
mean emission measure is important for testing proposals for possible
heating mechanisms. If the plasma is essentially in ionisation
equilibrium and the radiative cooling and (or) conductive terms
dominate(s) the energy balance, we expect the DEM to have everywhere
the same shape, differing only by a roughly constant factor depending
on the total amount of heating in that region. In this case, however,
the distribution of the plasma in temperature depends mainly on its
radiative and conductive properties and not on how and where the
energy release has taken place. On the other hand, if the radiative
cooling and (or) conduction are (is) not the dominant term in the
energy balance, then the DEM will depend on how and where the heating
has taken place and different structures may combine into a complex
mean DEM distribution, such as one with multiple peaks. Furthermore,
if the heating is impulsive, the plasma may not be in ionisation
equilibrium and analysis methods based on emission measure become
unreliable.

In this work we exploit the DEM in the spectral analysis of a quiet
Sun area observed by SOHO-CDS and address the issue of how the DEMs of
small subareas combine to give the DEM of a larger area. The
relationship with bulk motion is also investigated.
In Sect.\,\ref{sec:observations} we give an overview of the
observational data used in this work. In
Sect.\,\ref{sec:data_handling} we outline the reduction of such
data. In Sect.\,\ref{sec:DEM_analysis} we discuss the DEM analysis
performed on the observational data. We draw our conclusions in
Sect.\,\ref{sec:conclusions}.

\section{Overview of the CDS observing sequence}
\label{sec:observations}

\begin{figure}[t]
\hbox to \hsize {\hss \epsfxsize=88mm \epsfbox {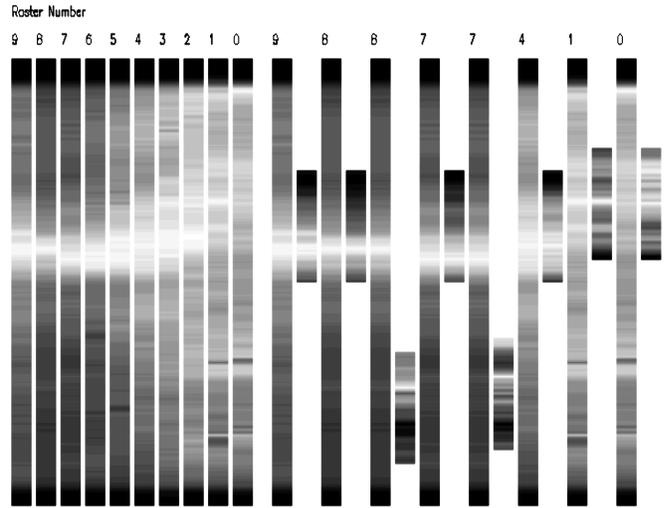}\hss}
   \caption{Same as Fig.\,\ref{fig:OV_raster} for
   \ion{Mg}{ix}\,(368.057\,\AA).}
   \label{fig:MGIX_raster}
\end{figure}

A full description of CDS scientific objectives and flight hardware is
given by \citet{Harrison_etal:95}. Details pertinent to the observing
sequence analysed in this paper are given in \citet{Brooks_etal:99}.

The data used for this work were taken using the NISAT\_S (ID 17
var.\,2) observing sequence. This takes the full wavelength ranges
from both the NIS channels, NIS-1 which covers the wavelength range
308-381\,\AA\ and NIS-2 which covers the range 513-633\,\AA. The 2
arcsec by 240 arcsec slit is used. This is the narrowest slit
available, yielding the best spectral resolution. The sequence raster,
covering an area of the Sun 20 arcsec by 240 arcsec, is made by moving
the scan mirror to 10 adjacent locations each separated by 2.032
arcsec (less a contribution from solar rotation). The exposure time at
each location is 50 s and the whole study takes 60 minutes to
complete, including data transmission time from the spacecraft.

The observations used in this paper were made on 1996 December 5 at
06\,h 58\,m (UT) as CDS sequence s6011r00 which was also one of the
datasets used by \citet{Brooks_etal:99}.  We therefore take full
advantage of the very detailed spectral analysis already carried out
for this dataset.  The pointing chosen was to the quiet Sun near Sun
centre \citep[see][]{Brooks_etal:99}.

\section{Data handling}
\label{sec:data_handling}

\begin{figure}[t]
\hbox to \hsize {\hss \epsfxsize=88mm \epsfbox {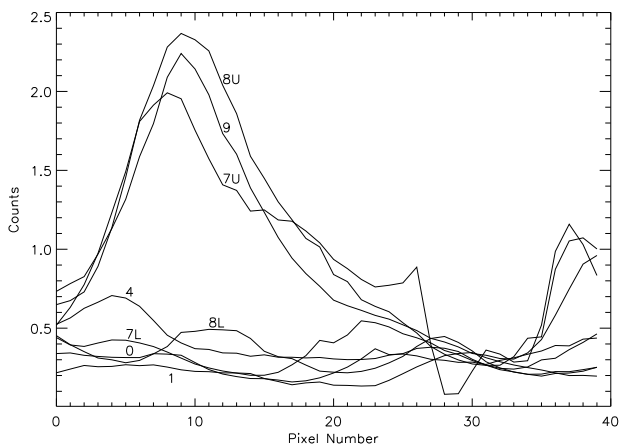}\hss}
   \caption{Variation of intensity in the \ion{O}{v}\,(629.732\,\AA)
   line along the slit for the 40 pixels long segments chosen.}
   \label{fig:OV_variation}
\end{figure}

\subsection{Reduction of raw data}

NIS detector data require debias correction, spectral rotation and
line tilt correction, cosmic ray strike cleaning, drop-out detection,
flat-fielding, correction for the localised ``burn-in'' of the
detector by strong lines and correction for non-linearity in detector
response. A discussion on the corrections applied to the raw data can
be found in \citet{Brooks_etal:99} \citep[see
also][]{Lang_etal:02}. The reduction procedure has been reapplied to
the raw data to take account of improved narrow slit burn-in and the
wide slit burn-in correction \citep{Lang_etal:02}. The spectral
observations were fitted to Gaussian profiles by the program ADAS602
as described in \citet{Brooks_etal:99}.

\subsection{Selection of subareas}
\label{sec:selection_of_subareas}

In this study we sought to contrast the DEM from subregions of a
raster with the DEM of the whole raster. To do so, we have chosen
eight segments, each 40 pixels long in the y-direction (i.e. along the
slit), in six of the exposures of the whole
raster. Figs.\,\ref{fig:OV_raster} and \ref{fig:MGIX_raster} show the
\ion{O}{v}\,(629.732\,\AA) and \ion{Mg}{ix}\,(368.057\,\AA) intensity
variation in the ten raster steps and then the position of the chosen
segment of 40 pixels along all the pixels in its step. We have chosen
some dark \ion{Mg}{ix}\,(368.057\,\AA) regions (4(7L) and 6(8L)),
regions around a narrow brightening (3(4), 5(7U), 7(8U) and 8(9)) and
a more extended brightening (1(0) and 2(1)).

\begin{figure}[t]
\hbox to \hsize {\hss \epsfxsize=88mm \epsfbox {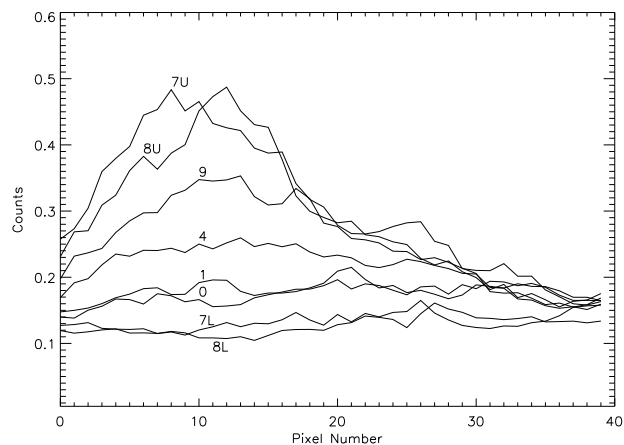}\hss}
   \caption{Same as Fig.\,\ref{fig:OV_variation} for
   \ion{Mg}{ix}\,(368.057\,\AA).}
  \label{fig:MGIX_variation}
\end{figure}

Figs.\,\ref{fig:OV_variation} and \ref{fig:MGIX_variation} show the
variation of the intensities of the \ion{O}{v}\,(629.732\,\AA) and
\ion{Mg}{ix}\,(368.057\,\AA) lines along the slit. The dark regions
show a rather flat intensity distribution along the slit, the extended
bright regions a flattish distribution with higher counts than in the
dark region, and the narrow brightening a peaked distribution with
much higher counts.

\subsection{Radiometric calibration}
\label{sec:rad_cal}

After fitting to Gaussian profiles, the radiometric calibration has
been applied to the spectral observations following
\citet{Lang_etal:02} and references therein.

To the standard error in flux returned by the Gaussian fitting
procedure ADAS602 \citep{Brooks_etal:99} we add, quadratically,
$\pm 8\,\%$ for aperture uncertainty, 
$\pm 2\,\%$ for slit size uncertainty and 
$\pm 10\,\%$ for burn-in correction for the line.
Also added in the quadrature to the estimated standard error were the
wavelength dependent uncertainties for the count/photon conversion.
That is for NIS-1 $\pm 15\,\%$ below 368\,\AA\ and $\pm 25\,\%$ above
368\,\AA, for NIS-2, $\pm 18\,\%$ at 584\,\AA\ and $\pm 29\,\%$ at the
ends of the band using linear interpolation between 584\,\AA\ and the
ends of the band and for the NIS-2 2nd order, $\pm 12\,\%$ at
303-304\,\AA.

%
\section{DEM analysis} 
\label{sec:DEM_analysis}
%

\subsection{Method}
\label{sec:DEM_method}

The method used to derive the differential emission measure, the
relevant part of the collisional-raditive theory for the line
emissivity, most of the atomic data used and the evaluation of kernels
have been discussed in detail in Paper I.  To the atomic data used in
Paper I, data for \ion{Mg}{iv}, \ion{Al}{vii}, \ion{Ar}{vii}, and
\ion{S}{v} have been added from the CHIANTI 4 database
(\citealt{Dere_etal:97}, \citealt{Young_etal:03}), and data for
\ion{Ca}{vii} and \ion{Ca}{viii} have been taken from
\citet{Landi_Bhatia:03} and \citet{Landi_etal:04}, respectively.

The lines used in the integral inversion and in the evaluation of
elemental abundances are indicated in Tables\,\ref{dem_line_list} and
5 with ``i'' (used in the integral inversion and evaluation of
abundance) and ``f'' (compared with observations in the forward sense
and used for the evaluation of elemental abundances). The selection of
these lines follows from the considerations made in
Sect.\,\ref{sec:line_analysis}.  
Table\,5, available at the CDS, contains the following
information. 
Columns 1 and 2 list the element symbol and the ion
charge ($z+1$), 
column 3 the adopted laboratory wavelength ($\lambda^{\rm ad}$), 
column 4 the reference for $\lambda^{\rm ad}$,
columns 5-8 the electronic configurations and terms,
column 9 the peak temperature of line formation $\log\,T_{\rm p}$,
column 10 the observed wavelength, 
column 11 the theoretical intensity,
column 12-13 the observed intensity and the observational uncertainty, 
column 14 the chi-square $(\sigma_0 \chi_i)^2$,
column 15 a note indicating whether the line has been used for the
integral inversion ``i'' or compared in the forward sense ``f''.
Intensities in erg cm$^{-2}$ s$^{-1}$ sr$^{-1}$ (mW m$^{-2}$ sr$^{-1}$
in SI units).

\begin{table*}[t]
\caption[]{Lines used for the integral inversion and for evaluating
           the elemental abundances. $\lambda^{\rm ad}$ is our
           preferred laboratory wavelength and $\lambda^{\rm cor}$ the
           ``corrected'' observed wavelength in
           \citet{Brooks_etal:99}.  $T_{\rm p}$ is the peak
           temperature of line formation. Since one line per ion at
           most is used in the integral inversion, we indicate with
           ``i'' the lines used in the integral inversion and ``f''
           the lines compared with observations in the forward sense,
           i.e. by comparing the intensities predicted using the DEM
           with the observed intensities. The source for the adopted
           wavelengthis is denoted by E for \citet{Edlen:83a,
           Edlen:83b, Edlen:84, Edlen:85a, Edlen:85b, Edlen:85c}, F
           for \citet{Fawcett:75}, K for \citet{Kelly:87}, N for
           \citet{Martin_etal:95}, and O for \citet{ORNL:85},
           respectively.}
\label{dem_line_list}
\begin{center}
\begin{tabular}{llrcllrc}
\hline
Ion & \multicolumn{1}{c}{$\lambda^{\rm ad}$} & 
\multicolumn{3}{c}{Transition} &
\multicolumn{1}{c}{$\lambda^{\rm cor}$}& $\log\,T_{\rm p}$ 
 & i/f \\
\hline
\ion{O}{iii}  & 525.797$^{\rm N}$ &   2s$^2$ 2p$^2$ $^1$D$_{2}$ & - & 2s 2p$^3$ $^1$P$_{1}$       & 525.801 & 4.95 & f \\ 
\ion{O}{iii}  & 599.598$^{\rm N}$ &   2s$^2$ 2p$^2$ $^1$D$_{2}$ & - & 2s 2p$^3$ $^1$D$_{2}$       & 599.596 & 5.00 & i \\
\ion{O}{iv}   & 553.329$^{\rm N}$ &      2s$^2$ 2p$^2$P$_{1/2}$ & - & 2s 2p$^2$ $^2$P$_{3/2}$     & 553.343 & 5.20 & f \\
\ion{O}{iv}   & 554.076$^{\rm N}$ &     2s$^2$ 2p $^2$P$_{1/2}$ & - & 2s 2p$^2$ $^2$P$_{1/2}$     & 554.076 & 5.20 & f \\
\ion{O}{iv}   & 554.513$^{\rm N}$ &     2s$^2$ 2p $^2$P$_{3/2}$ & - & 2s 2p$^2$ $^2$P$_{3/2}$     & 554.512 & 5.20 & f \\
\ion{O}{iv}   & 555.263$^{\rm N}$ &     2s$^2$ 2p $^2$P$_{3/2}$ & - & 2s 2p$^2$ $^2$P$_{1/2}$     & 555.270 & 5.20 & i \\
\ion{O}{iv}   & 608.397$^{\rm N}$ &     2s$^2$ 2p $^2$P$_{1/2}$ & - & 2s 2p$^2$ $^2$S$_{1/2}$     & 608.312 & 5.20 & f \\
\ion{Ne}{iv}  & 543.886$^{\rm E}$ & 2s$^2$ 2p$^3$ $^4$S$_{3/2}$ & - & 2s 2p$^4$ $^4$P$_{5/2}$     & 543.881 & 5.30 & i \\
\ion{O}{v}    & 629.732$^{\rm N}$ &          2s$^2$ $^1$S$_{0}$ & - & 2s 2p $^1$P$_{1}$           & 629.735 & 5.40 & i \\
\ion{Ar}{vii} & 585.754$^{\rm F}$ &          3s$^2$ $^1$S$_{0}$ & - & 3s 3p $^1$P$_{1}$           & 585.694 & 5.50 & f \\
\ion{Ne}{v}   & 359.375$^{\rm E}$ &     2s$^2$ 2p$^2$ $^3$P$_2$ & - & 2s 2p$^3$ $^3$S$_1$         & 359.374 & 5.50 & i \\
\ion{Ne}{vii} & 561.72$^{\rm E}$  &             2s 2p $^3$P$_2$ & - & 2p$^2$ $^3$P$_2$            & 561.725 & 5.70 & i \\
\ion{Al}{vii} & 356.892$^{\rm E}$ & 2s$^2$ 2p$^3$ $^4$S$_{3/2}$ & - & 2s 2p$^4$ $^4$P$_{5/2}$     & 356.922 & 5.75 & f \\
\ion{Ca}{viii}& 582.845$^{\rm K}$ &     3s$^2$ 3p $^2$P$_{1/2}$ & - & 3s 3p$^2$ $^2$D$_{3/2}$     & 582.872 & 5.75 & f \\
\ion{Ca}{x}   & 557.759$^{\rm E}$ &            3s $^2$S$_{1/2}$ & - & 3p $^2$P$_{3/2}$            & 557.764 & 5.80 & i \\
\ion{Al}{viii}& 328.183$^{\rm E}$ &     2s$^2$ 2p$^2$ $^3$P$_2$ & - & 2s 2p$^3$ $^3$P$_2$         & 328.249 & 5.90 & f \\
\ion{Al}{viii}& 328.230$^{\rm E}$ &    2s$^2$  2p$^2$ $^3$P$_2$ & - & 2s 2p$^3$ $^3$P$_1$         &         & 5.90 & f \\
\ion{Si}{viii}& 316.218$^{\rm E}$ & 2s$^2$ 2p$^3$ $^4$S$_{3/2}$ & - & 2s 2p$^4$ $^4$P$_{3/2}$     & 316.215 & 5.90 & f \\
\ion{Si}{viii}& 319.839$^{\rm E}$ & 2s$^2$ 2p$^3$ $^4$S$_{3/2}$ & - & 2s 2p$^4$ $^4$P$_{5/2}$     & 319.840 & 5.90 & f \\
\ion{Mg}{viii}& 311.772$^{\rm E}$ &     2s$^2$ 2p $^2$P$_{1/2}$ & - & 2s 2p$^2$ $^2$P$_{3/2}$     & 311.750 & 5.90 & f \\
\ion{Mg}{viii}& 313.743$^{\rm E}$ &     2s$^2$ 2p $^2$P$_{1/2}$ & - & 2s 2p$^2$ $^2$P$_{1/2}$     & 313.758 & 5.90 & i \\
\ion{Mg}{viii}& 335.231$^{\rm E}$ &     2s$^2$ 2p $^2$P$_{1/2}$ & - & 2s 2p$^2$ $^2$S$_{1/2}$     & 335.233 & 5.90 & f \\
\ion{Fe}{xi}  & 352.661$^{\rm O}$ &     3s$^2$ 3p$^4$ $^3$P$_2$ & - & 3s 3p$^5$ $^3$P$_2$         & 352.671 & 6.05 & f \\
\ion{Si}{ix}  & 341.950$^{\rm E}$ &     2s$^2$ 2p$^2$ $^3$P$_0$ & - & 2s 2p$^3$ $^3$D$_1$         & 341.952 & 6.05 & i \\
\ion{Al}{x}   & 332.788$^{\rm E}$ &            2s$^2$ $^1$S$_0$ & - & $2s2p$ $^1$P$_1$            & 332.785 & 6.10 & f \\
\ion{Al}{xi}  & 550.04$^{\rm E}$  &            2s $^2$S$_{1/2}$ & - & 2p $^2$P$_{3/2}$            & 550.050 & 6.15 & f \\
\ion{Si}{x}   & 347.408$^{\rm E}$ &     2s$^2$ 2p $^2$P$_{1/2}$ & - & 2s 2p$^2$ $^2$D$_{3/2}$     & 347.402 & 6.15 & i \\
\ion{Fe}{xii} & 364.468$^{\rm O}$ & 3s$^2$ 3p$^3$ $^4$S$_{3/2}$ & - & 3s 3p$^4$ $^4$P$_{5/2}$     & 364.467 & 6.15 & f \\
\ion{Si}{xii} & 520.662$^{\rm E}$ &            2s $^2$S$_{1/2}$ & - &2p $^2$P$_{1/2}$             & 520.685 & 6.30 & i \\
\hline
\end{tabular}
\end{center}
\end{table*}

The DEM is evaluated separately for the mean spectrum and for each
segment, adopting a constant electron density model for the evaluation
of kernels as in Tables\,\ref{table:MgVIII_line_ratios} and
\ref{table:segments_Ne}. A constant pressure model has also been
considered, but, since the results are not appreciably different, this
is not reported here.
Anticipating the results of Sect.\,\ref{sec:B-like},
Table\,\ref{table:MgVIII_line_ratios} shows that the electron density
in the mean spectrum at the formation temperature of \ion{Mg}{viii} is
approx. $\log\,N_{\rm e}\,=\,7.7$. In addition,
Table\,\ref{table:segments_Ne} shows that the electron density in the
8 segments varies from $\log\,N_{\rm e}\,=\,6.9$ to 8.3.

The evaluation of elemental abundances is performed independently for
each segment and for the mean spectrum (see
Sect.\,\ref{sec:abundance}).

\subsection{Spectral analysis}
\label{sec:line_analysis}

A preliminary analysis using line intensity ratios is performed for
selecting, for the integral inversion, lines not significantly
affected by blending, and to infer the most likely electron density to
be used in the evaluation of the $G$-functions. The selection is made
with the aim of reducing, as much as possible, systematic errors due
to the ubiquitous line blending. The spectral analysis of
\citet{Brooks_etal:99} is used as a starting point for lines and
blends identification. Here we consider mainly lines to which I-codes
A and B and P-codes 1 and 2 were assigned and which are neither
blended with lines of other ions nor affected by opacity. In the
following, we indicate with $R_{\rm th}$ the theoretical line
intensity ratio obtained with ADAS. The spectral analysis is then
complemented with the DEM fit, which allows a detailed study of blends
of different ions and a check of the results obtained from the line
ratio analysis.
For the aid of the reader in understanding why some of the lines were
rejected from the DEM analysis in the subsequent sections, we note
that the final DEM solution was found to be most reliable in the
temperature range $\log\,T_{\rm e}\,=\,5.0$ to 6.0. The results are given
in Table\,5 (available at CDS).

\subsubsection{Li-like sequence}
\label{sec:Li-like}

Both lines of the \ion{Mg}{x} 2s\,$^2$S - 2p\,$^2$P doublet
(624.95\,\AA, 609.79\,\AA) were identified by \citet{Brooks_etal:99},
the 1/2-1/2 component being of code A[1] and the 1/2-3/2 component
being blended with \ion{O}{iv} which \citeauthor{Brooks_etal:99}
estimated, from branching ratios, to contribute about 1/3 of the
intensity of the blend.  Good agreement is found with the DEM
reconstruction in the mean spectrum. However, the DEM fails to
reproduce the observed intensities in some segments, as shown in
Table\,\ref{table:Mgix-x_comparison}.  This can be due to several
reasons, including inaccuracies in the atomic data, a complex
distribution of the plasma in electron density ($N_{\rm e}$), or to
the DEM uncertainties which increase with temperature above
$\log\,T\,=\,6.0$.  Another possible explanation could be that some
uncertainties in the NIS-1/NIS-2 cross-calibration still exist, but
this is ruled out by the considerations made in
Sect.\,\ref{sec:cross-calibration}.  Note, however, that when
\ion{Mg}{x} is reproduced by the DEM reconstruction, \ion{Mg}{ix}
(Be-like isoelectronic sequence, see next Sect. and
Table\,\ref{table:Mgix-x_comparison}) is not and vice
versa. Furthermore, $I_{\rm th}/I_{\rm obs}$ for the \ion{Mg}{x} lines
is systematically higher than for the \ion{Mg}{ix} line.  As discussed
in Sect.\,\ref{sec:Be-like}, the comparison with lines of \ion{Mg}{ix}
(Be-like) suggests that such discrepancies are likely to be caused by
either some degree of non-equilibrium in the ionisation balance or a
problem with the ionisation balance, e.g. by a recombination
contribution to the 2s 2p $^1$P term of \ion{Mg}{ix} or ionisation to
it from e.g. the 2s 2p$^2$ $^4$P term in the B-like \ion{Mg}{viii},
noting that as \ion{Mg}{viii} is used to construct the DEM, problems
in that ion will not be identified.

The same doublet was identified for \ion{Al}{xi} by
\citet{Brooks_etal:99}, the 1/2-3/2 component at 550.04\,\AA\ and the
1/2-1/2 component at 568.16\,\AA\ blended with \ion{Ne}{v}. The DEM
reconstruction reproduces fairly well \ion{Al}{xi}\,(550.04\,\AA) and
predicts a 10\,\% contribution to the blend with \ion{Ne}{v} at
568.16\,\AA. Note, however, the large uncertainties in the DEM at the
formation temperature of this line and in the \element{Al} abundance
(see Sect.\,\ref{sec:abundance}).  Nevertheless, $I_{\rm th}/I_{\rm
obs}$ for the \ion{Al}{xi}\,(550.04\,\AA) line is found to be
systematically higher than for the \ion{Al}{x}\,(332.788\,\AA) line
(Be-like, see next Sect.), which indicates the same systematic
behavior found for the \ion{Mg}{x} and \ion{Mg}{ix} lines.

\subsubsection{Be-like sequence}
\label{sec:Be-like}

\ion{O}{v}\,(629.732\,\AA) is a strong NIS-2 calibration line and is
used here both for selecting the subareas
(Sect.\,\ref{sec:selection_of_subareas}) and for the integral
inversion to sample the region around $\log\,T\,=\,5.4$. 

\citet{Brooks_etal:99} identified the 2s\,2p\,$^3$P - 2s\,3s\,$^3$S
transitions in \ion{C}{iii}, the 0-1 (538.080\,\AA) and the 1-1
(538.149\,\AA) blended together and with \ion{O}{ii}\,(537.832\,\AA),
and the 2-1 (538.312\,\AA) blended with \ion{O}{ii}\,(538.262\,\AA)
and \ion{O}{ii}\,(538.320\,\AA). Such blends may be affected by line
opacities and their formation temperature is below the trusted
temperature range for the DEM reconstruction ($\log\,T \sim 4.75$ for
\ion{O}{ii} and $\log\,T \sim 4.95$ for \ion{C}{iii}). The results for
these lines are therefore only indicative and are reported in
Tab.\,5.

The \ion{Ne}{vii} 2s\,2p\,$^3$P - 2p$^2$\,$^3$P multiplet has two
lines known to be blended with \ion{Ne}{vi} (1-2 at 558.60\,\AA\ and
1-0 at 562.98\,\AA). The other four have large error bars because of
their low intensity, which does not allow a reliable line ratio
analysis.  The 561.38/559.95\,\AA\ ratio in the mean spectrum is 0.94
which, when compared with $R_{\rm th}\,=\,0.75$, may indicate a blend
affecting the 561.38\,\AA\ line, although the large observational
error bars do not allow a definitive conclusion.  A similar
consideration can be made for the 564.52\,\AA\ line, since the
564.52/559.95\,\AA\ ratio is 1.54 in the mean spectrum while $R_{\rm
th}\,=\,1.22$.  The \ion{Ne}{vii} 2s\,2p\,$^1$P$_1$ -
2p$^2$\,$^1$S$_0$ line at 561.27\,\AA\ could contribute to a blend
with 561.38\,\AA; in this case the composite line ratio becomes
essentially unity above $N_{\rm e}\,=\,10^8$ cm$^{-3}$ and shows a
weak density dependence below such value, rising to 1.7 at $N_{\rm
e}\,=\,10^6$ cm$^{-3}$.  Density sensitive ratios indicate only a
lower limit to the electron density, which must be above
$\log\,N_e\,=\,7$ to obtain line ratios consistent with the
observations. Despite such uncertainties, we are forced to select, for
the integral inversion, the strongest line apparently not affected by
significant blending (2-2 at 561.72\,\AA) because of the lack of lines
sampling the temperature region around $\log\,T_{\rm e}\,=\,5.7$.  The
DEM analysis does not confirm the blending of the lines at
561.38\,\AA\ and 564.52\,\AA, which are reproduced within the
observational uncertainties, although the results are also compatible
with the blending of the 561.38\,\AA\ and 561.27\,\AA\ lines. It
indicates, instead, further blending for the lines at 558.60\,\AA\ and
562.98\,\AA\ or a problem in reproducing the \ion{Ne}{vi} line
intensities.

\begin{table*}[t]
\caption[]{Comparison between the
           Be-like \ion{Mg}{ix}\,(368.057\,\AA) and the Li-like 
           \ion{Mg}{x}\,(624.95\,\AA) transitions in the selected segments.}
\label{table:Mgix-x_comparison}
\begin{center}
\begin{tabular}{llccccl}
\hline
 &    &  \multicolumn{2}{c}{Li-like}                    & \multicolumn{2}{c}{Be-like}                      &  \\
 &    &  \multicolumn{2}{c}{\ion{Mg}{x}\,(624.95\,\AA)} & \multicolumn{2}{c}{\ion{Mg}{ix}\,(368.057\,\AA)} & \\
Regions & segment & $I_{\rm th}/I_{\rm obs}$ & $(\sigma_0 \chi_i)^2$ & $I_{\rm th}/I_{\rm obs}$ & $(\sigma_0 \chi_i)^2$ & note \\
\hline
Extended Brightening & 1(0)  & 1.1 & 1.3(-1) & 0.5  & 3.4     & \\
                     & 2(1)  & 0.6 & 1.8     & 0.2  & 7.2     & high velocities \\
\\
Narrow brightening   & 3(4)  & 0.7 & 1.3     & 0.4  & 4.8     & low intensities \\
                     & 5(7U) & 1.9 & 8.2     & 0.8  & 2.8(-1) & \\
                     & 7(8U) & 1.1 & 7.1(-1) & 0.5  & 3.3     & \\
                     & 8(9)  & 2.5 & 2.5(+1) & 1.0  & 1.4(-2) & \\
\\
Dark regions         & 4(7L) & 0.5 & 2.3     & 0.3  & 6.4     & high velocities \\
                     & 6(8L) & 1.0 & 1.2(-4) & 0.5  & 4.8     & high velocities \\
\hline
\end{tabular}
\end{center}
\end{table*}

For the resonance line \ion{Mg}{ix}\,(368.057\,\AA) $I_{\rm th}$ is
about half $I_{\rm obs}$ in the mean spectrum.  However, this line is
reproduced within the observational uncertainties in segments 5(7U),
7(8U), and 8(9), i.e. in the segments around a narrow brightening with
higher intensities than average.  $I_{\rm th}$/$I_{\rm obs}$ is about
0.2 -- 0.5 in the remaining segments (see
Table\,\ref{table:Mgix-x_comparison}).  This may be due to a blend
that occasionally affects this line, uncertainties in the extra
burn-in correction for strong lines or DEM uncertainties which
increase with temperature above $\log\,T\,=\,6.0$.  Uncertainties in
the collisional excitation coefficients can obviously be another cause
of the discrepancies.  Note, however, that the situation is the
opposite to that in the Li-like \ion{Mg}{x} lines
(Sect.\,\ref{sec:Li-like}).  In fact, comparing the Li-like
\ion{Mg}{x}\,(624.95\,\AA) with the Be-like
\ion{Mg}{ix}\,(368.057\,\AA) in each segment we note that the ratio
$I_{\rm th}/I_{\rm obs}$ for the former is systematically higher than
for the latter.  This situation weakens the explanation of the
discrepancies as due to occasional blends or uncertainties in the
calibration of the instrument. Recombination to the 2s 2p $^1$P term
or ionisation to it from e.g. the 2s 2p$^2$ $^4$P term in the B-like
\ion{Mg}{viii} could increase its population density and therefore
increase the \ion{Mg}{ix}\,(368.057\,\AA) intensity. Since
\ion{Mg}{viii} is used to construct the DEM, problems in that ion will
not be identified. However, level-resolved ionisation and
recombination coefficients for \element{Mg} are not yet available and
therefore this possibility cannot be checked. Such a systematic
behaviour can also be interpreted as being due to an {\it
overionisation} of the Li-like stage with respect to the equilibrium
ionisation balance. This behaviour could appear evident in our
analysis because of the use of \element{Si} lines in the calculation
of the DEM at the formation temperatures of these lines
($\log\,T\,\sim\,6.0$ and above), which have relaxation time-scales
for the ground states about 50\,\% shorter than \element{Mg} at such
temperatures (see Paper I).  \ion{Mg}{ix}\,(368.057\,\AA) was used in
the integral inversion for the SERTS-89 AR spectrum (Paper I) and
reproduced within the observational uncertainties.

The \ion{Al}{x} 2s$^2$ $^1$S$_0$ - 2s 2p $^1$P$_1$ transition at
332.788\,\AA\ was used in the integral inversion in Paper I and was
found to be in excellent agreement with observation.
\citet{Brooks_etal:99} assigned code B[2] to this line since its
position pattern was found not to be as consistent with temperature as
expected.  The DEM predicts an intensity much lower than
observed. Note, however, the large uncertainties in the DEM at the
formation temperature of this line and in the \element{Al} abundance
(see Sect.\,\ref{sec:abundance}).  As noted in
Sect.\,\ref{sec:Li-like}, $I_{\rm th}/I_{\rm obs}$ for the
\ion{Al}{xi}\,(550.04\,\AA) line (Li-like) is found to be
systematically higher than for the \ion{Al}{x}\,(332.788\,\AA)
line. This supports the conclusion drawn from the comparison between
the Li-like \ion{Mg}{x}\,(624.95\,\AA) and the Be-like
\ion{Mg}{ix}\,(368.057\,\AA) lines of an overionisation of the Li-like
stage with respect to the equilibrium ionisation balance, but, as
outlined for \ion{Mg}{ix}, level-resolved ionisation balance
calculations are needed to confirm this conclusion.

Of the \ion{Si}{xi} transitions, the 2s$^2$ $^1$S$_{0}$ - 2s 2p
$^1$P$_{1}$ at 303.326\,\AA\ was identified by \citet{Brooks_etal:99}
in NIS-2 2$^{\rm nd}$ order. Other \ion{Si}{xi} lines analysed in
Paper I are expected to be below the sensitivity of the instrument in
a QS spectrum.  The DEM reconstruction for the 2s$^2$ $^1$S$_{0}$ --
2s 2p $^3$P$_{1}$ (580.92\,\AA) intercombination line yields a value
0.2 of the observed.  \citet{Brooks_etal:99} assigned I-code E to
\ion{Si}{xi}\,(604.12\,\AA) (2s\,2p\,~$^1$P$_{1}$ -
2p$^2$\,$^1$D$_{2}$) because of its large wavelength difference with
the $\lambda^{\rm ad}$. Here it is predicted to have an intensity of
only 0.06 erg cm$^{-2}$ s$^{-1}$ sr$^{-1}$ in the mean spectrum and
therefore its identification in the QS in NIS-2 is rejected.  For
\ion{Si}{xi}\,(303.326\,\AA) we find $I_{\rm th}/I_{\rm obs}\sim 0.1$
in the mean spectrum and between 0.1 and 0.6 in the selected segments;
the theoretical intensity is closer to observation (within 1.5
$\sigma$) in segments 1(0), 2(1), 4(7L), 5(7U), and 6(8L).  This
discrepancy is consistent with the SERTS-89 spectral analysis of Paper
I, where this line had $I_{\rm th}/I_{\rm obs}\sim 0.5$, and may at
least partially be ascribed to the uncertainties of the QS DEM above
$\log\,T_{\rm e}\,=\,6.0$.  Possible blends are
\ion{Fe}{xiii}\,(303.320\,\AA) and \ion{S}{xiii}\,(303.384\,\AA), but
neither could contribute appreciably to the observed intensity. Note
that this line may suffer from radiative pumping because of its close
vicinity to the strong \ion{He}{ii}\,(303.78\,\AA).  Possible
uncertainties in the atomic data have been discussed by
\citet{Lang_etal:01}. A more robust analysis of
\ion{Si}{xi}\,(303.326\,\AA) in NIS spectra could be done considering
AR observations for which the DEM above $\log\,T_{\rm e}\,=\,6.0$ is
expected to be more reliable.

\begin{table}[t]
\caption[]{Electron density derived from \ion{Mg}{viii} lines ratios.
The 315.015\,\AA\ and 335.231\,\AA\ line intensities have been
corrected assuming that they are affected by blendings contributing
50\,\% and 20\,\% of the observed intensity, respectively. $C$ is the
correction factor applied to the line ratio to take blending into
account.}
\label{table:MgVIII_line_ratios}
\begin{tabular}{lll}
\hline
Ratio (\AA)      & $\log\,N_{\rm e}$   & $C$ \\
\hline
311.772/313.743  & 7.7$_{-0.3}^{+0.3}$ & 1.00 \\
315.015/313.743  & 7.8$_{-0.2}^{+0.3}$ & 0.67 \\
335.231/313.743  & 7.8$_{-0.5}^{+0.5}$ & 0.82 \\
338.983/313.743  & 7.8$_{-0.5}^{+0.5}$ & 1.00 \\
317.028/315.015  & 7.7$_{-0.2}^{+0.3}$ & 1.50 \\
335.231/315.015  & 7.7$_{-0.4}^{+0.7}$ & 1.24 \\
338.983/315.015  & 7.8$_{-0.4}^{+0.9}$ & 1.50\\
311.772/317.028  & 7.7$_{-0.3}^{+0.3}$ & 1.00 \\
335.231/317.028  & 7.7$_{-0.6}^{+0.5}$ & 0.82 \\
338.983/317.028  & 7.7$_{-0.4}^{+0.6}$ & 1.00 \\
\hline
\end{tabular}
\end{table}

\subsubsection{B-like sequence}
\label{sec:B-like}

The observed \ion{O}{iv} line ratios are not sensitive to
density. Lines indicated as unblended by \citet{Brooks_etal:99} form
ratios in agreement with theory. The strongest line at 554.513\,\AA\
is selected for the integral inversion.  \ion{O}{iv}\,(609.829\,\AA)
is blended with \ion{Mg}{x} and, as discussed in
Sect.\,\ref{sec:Be-like}, the DEM reconstruction is in good agreement
with observation. The 5/2-3/2, 3/2-3/2, and 3/2-1/2 components of the
2s 2p$^2$ $^2$D - 2p$^3$ $^2$P multiplet (616.952\,\AA, 617.005\,\AA,
617.036\,\AA) are reported as blended with \ion{O}{ii}. The
discrepancies observed may be due to our difficulties in reproducing
the \ion{O}{ii} lines, either because of opacity effects and/or
because this line is formed below our trusted range of temperatures
for the evaluation of the DEM.

All the 2s$^2$\,2p\,$^2$P - 2s\,2p$^2$\,$^2$S and 2s$^2$\,2p\,$^2$P -
2s\,2p$^2$\,$^2$P transitions in \ion{Mg}{viii} were classified as
code B[1] by \citet{Brooks_etal:99}. The intensity ratio of lines
sharing the same upper level shows that the 317.028/313.743\,\AA\
ratio is in excellent agreement with the theory ($R_{\rm
th}\,=\,0.64$).  The 311.772/315.015\,\AA\ ($R_{\rm th}\,=\,0.20$) and
338.983/335.231\,\AA\ ($R_{\rm th}\,=\,1.26$) ratios, despite their
theoretical values lying within the observational error bars, show
systematic departures in the average spectrum and in the segments that
may indicate that the 315.015\,\AA\ and 335.231\,\AA\ lines are
affected by blends contributing 50\,\% and 20\,\% of the observed
intensity, respectively. Correcting the average intensities using
these estimates, we obtain a remarkable agreement in the electron
density estimated by all the density sensitive \ion{Mg}{viii} line
ratios (see Table\,\ref{table:MgVIII_line_ratios}).
\ion{Mg}{viii}\,(311.772\,\AA) was reported as blended with
\ion{Ni}{xv} in Paper I, but this is expected to have negligible
intensity in a QS spectrum.  \ion{Mg}{viii}\,(315.015\,\AA) was
selected for the integral inversion and reproduced within the
observational uncertainties in Paper I, which may indicate that the
blending is due to a line formed at temperatures below that of
\ion{Mg}{viii} formation and becomes negligible compared with
\ion{Mg}{viii}\,(315.015\,\AA) in AR
conditions. \ion{Mg}{viii}\,(335.231\,\AA) is completely masked by
\ion{Fe}{xvi}\,(335.396\,\AA) in the SERTS-89 spectrum and
\ion{Mg}{viii}\,(338.983\,\AA) was found in excellent agreement with
observations in Paper I. We select \ion{Mg}{viii}\,(313.743\,\AA) for
the integral inversion and the ratio 338.983/313.743\,\AA\ as density
diagnostic at the temperature peak of line formation 
($\log\,T_{\rm p}\,\approx\,5.90$\,K). 
The electron density in the selected segments, inferred from the
338.983/313.743\,\AA\ intensity ratio, is shown in
Table\,\ref{table:segments_Ne}.  Given the reliability of this density
diagnostic, such values are used for evaluating the $G$-functions. The
DEM analysis confirms the conclusions drawn from the line ratio
analysis. The 315.015\,\AA\ line is underestimated by 45\,\% in the
mean spectrum, lying outside the error bar, and systematically
underestimated in all the segments, with $I_{\rm th}/ I_{\rm obs}$
between 0.2 (segment 2(1)) and 0.9 (segment 8(9)).  The line at
335.231\,\AA\ is underestimated by 30\,\% in the mean spectrum, lying
outside the error bar, and systematically underestimated in all the
segments, with $I_{\rm th}/ I_{\rm obs}$ between 0.4 (segments 2(1)
and 4(7L)) and 0.8 (segment 5(7U) and 8(9)).  \citet{Brooks_etal:99}
noted the absence of the \ion{Fe}{xvi} 3s $^2$P - 3p $^2$P lines at
335.396\,\AA\ and 360.743\,\AA\ which are bright in active region
spectra (see also Paper I). The present analysis is consistent with
the stronger 335.396\,\AA\ line blending with the
\ion{Mg}{viii}\,(335.231\,\AA) line. \citet{Brooks_etal:99} also
indicated that the \ion{Fe}{xii}\,(335.06\,\AA) line may be lost in
the wings of the \ion{Mg}{viii}\,(335.231\,\AA) line. However, it
should be noted that \citet{Lang_etal:02} pointed out that for the
\ion{Fe}{xvi} doublet there could be extra wide-slit burn-in which is
not accommodated by the anlysis. This could affect the intensity of
the \ion{Mg}{viii}\,(335.231\,\AA) line.

\begin{table}
\caption[]{Electron density derived from the
          \ion{Mg}{viii} 338.983/313.743\,\AA\  
          line ratio in the selected segments.
          The lower limit and the upper limit are 
          indicated with ``ll'' and ``ul'', respectively.}
\label{table:segments_Ne}
\begin{tabular}{ll}
\hline
Segment & $\log\,N_{\rm e}$ \\
\hline
1(0)   & 7.3$_{\rm ll}^{+0.5}$ \\
2(1)   & 6.9$_{\rm ll}^{+0.6}$ \\
3(4)   & 7.4$_{\rm ll}^{+0.5}$ \\
4(7L)  & 8.0$_{-0.6}^{+1.7}$ \\
5(7U)  & 8.2$_{-0.6}^{\rm ul}$ \\
6(8L)  & 7.4$_{\rm ll}^{+0.5}$ \\
7(8U)  & 7.6$_{-0.9}^{+0.5}$ \\
8(9)   & 8.3$_{-0.6}^{\rm ul}$ \\
\hline
\end{tabular}
\end{table}

The \ion{Ne}{vi} 2s$^2$\,2p\,$^2$P - 2s\,2p$^2$\,$^2$D transitions lie
in the NIS-2 wavelength range. The 1/2-3/2 transition (558.591\,\AA)
is blended with \ion{Ne}{vii}. The 3/2-3/2 line (562.702\,\AA) is
assumed to be blended with the 3/2-5/2 line (562.797\,\AA) and with
\ion{Ne}{vii}\,(562.98\,\AA). No useful ratio can therefore be formed
with such lines. Table\,5 shows that both blends are dominated by the
\ion{Ne}{vi} components which, however, are overestimated by 70-90\,\%
in the DEM reconstruction. Note that in the SERTS-89 spectrum (Paper
I) the \ion{Ne}{vi} 2s$^2$ 2p $^2$P - 2s 2p$^2$ $^2$P multiplet was
reproduced within the observational uncertainties, but the 2s$^2$ 2p
$^2$P - 2s 2p$^2$ $^2$S multiplet was overestimated by a factor of
2. This situation points to inaccuracies in the atomic data as the
possible cause of the discrepancies.  We are, nevertheless, forced to
use this multiplet to constrain the \element{Mg}/\element{Ne}
abundance ratio (see Sect.\,\ref{sec:abundance}), obtaining consistent
results.

The \ion{Si}{x} 2s$^2$\,2p\,$^2$P - 2s\,2p$^2$\,$^2$D lines are
observed in the NIS-1 wavelength range, the 3/2-3/2 (356.054\,\AA) and
3/2-5/2 (356.029\,\AA) components being blended together. The
composite line ratio obtained with this multiplet indicates $N_{\rm e}
\approx 6 \times 10^{8}$ cm$^{-3}$, which is too high compared with
the value obtained from \ion{Mg}{viii}. The DEM analysis indicates
that further blending may affect these lines. However, they were {\it
overestimated} in the SERTS-89 analysis, which suggests inaccuracies
in the atomic data instead.  The 1/2-3/2 component at 347.408\,\AA,
which was in excellent agreement with observation in the SERTS-89
spectrum, is chosen for the integral inversion and is in good
agreement with observation.  These lines form in a temperature range
where large uncertainties affect the DEM. Note that the temperature of
line formation overlaps with that of \ion{Fe}{xii}.

\subsubsection{C-like sequence}
\label{sec:C-like}
%
\citet{Brooks_etal:99} assigned code A[1] to the 2s$^2$ 2p$^2$
$^1$D$_{2}$ - 2s 2p$^3$ $^1$D$_{2}$ transition in \ion{O}{iii} at
599.598\,\AA, and to the 2s$^2$ 2p$^2$ $^1$D$_{2}$ - 2s 2p$^3$
$^1$P$_{1}$ transition at 525.795\,\AA. The former is used here for
the integral inversion and the latter is compared with observation in
the forward sense. Both are reproduced within the observational
uncertainties. \ion{O}{iii} at 597.818\,\AA\ (2s$^2$ 2p$^2$
$^1$S$_{0}$ - 2s 2p$^3$ $^1$P$_{1}$) was reported as blended with
\ion{Ca}{viii}\,(597.851\,\AA) (3s$^2$ 3p $^2$P$_{3/2}$ - 3s 3p$^2$
$^2$D$_{3/2}$) by \citet{Brooks_etal:99}, who estimated a 10\,\%
contribution by \ion{Ca}{viii} to the blend. This is confirmed by the
DEM analysis which reproduces this blend within the observational
uncertainties.

The \ion{O}{iii} 2s$^2$ 2p$^2$ $^3$P - 2s$^2$ 2p 3s $^3$P transitions
were identified in NIS-1 by \citet{Brooks_etal:99} at $\lambda^{\rm
cor}$\,=\,373.804\,\AA\ (1-2 code C[3]), 374.097\,\AA\ (0-1, 2-2 and
1-1, code B[1]), and 374.433\,\AA\ (1-0 and 2-1, code B[2]). The
relative intensities of the lines were found in accord with
expectation and are density insensitive.  These lines were found in
good agreement with observations in the SERTS-89 spectrum (Paper
I). The DEM reconstruction in the mean CDS spectrum and in the
individual segments is, however, lower than observations ($I_{\rm
th}/I_{\rm obs} \sim 0.2$).  Given also the good agreement found with
\ion{O}{iii}\,(599.598\,\AA) and \ion{O}{iii}\,(525.795\,\AA), we
explored the possibility that such discrepancies may be due to
inaccuracies in the NIS-1/NIS-2 cross calibration.  We found, however,
that the correction required to bring these lines into agreement would
be well above the estimated uncertainties in the cross calibration
(see \citealt{Lang_etal:02} and Sect.\,\ref{sec:cross-calibration})
and therefore the reason for such discrepancy remains unclear.

The \ion{Ne}{v} 2s$^2$\,2p$^2$\,$^3$P - 2s\,2p$^3$\,$^3$D transitions
are observed in NIS-2 as a blend of the 2-3 and 2-2 components
(572.334\,\AA, 572.112\,\AA), a blend of the 1-1 and 1-2 components
(569.758\,\AA, 569.836\,\AA) and a blend of the 0-1 component
(568.421\,\AA) with \ion{Al}{xi}. The ratio formed with the 2-3 plus
2-2 composite assembly and the 1-1 plus 1-2 is density insensitive and
in agreement with theory. This multiplet is overestimated by the DEM
reconstruction by at least 60\,\%. A comparison with other \ion{Ne}{v}
lines points to uncertainties in the atomic data as a possible cause
for the discrepancies, but we note that the DEM at the formation
temperature of these lines is mainly constrained by the
\ion{Ne}{v}\,(359.375\,\AA) line whose choice is mainly supported by
the lack of other unblended lines in the temperature range rather than
by a line ratio analysis (see below). The fact that our analysis
overestimates these NIS-2 \ion{Ne}{v} lines while using a NIS-1
\ion{Ne}{v} line to constrain the DEM would point also, as discussed
above for \ion{O}{iii}, to a possible inaccuracy of the NIS-1/NIS-2
cross-calibration, but this is excluded as above.

Of the \ion{Ne}{v} 2s$^2$\,2p$^2$\,$^3$P - 2s\,2p$^3$\,$^3$S
transitions, observed in NIS-1, the 0-1 (357.946\,\AA) is thought to
be blended with \ion{Ne}{iv} and the 1-1 (358.476\,\AA) with
\ion{Fe}{x}.  The remaining line (2-1 at 359.375\,\AA) was classified
as code B[2] and was used in the wavelength calibration. This line is
also selected here for the integral inversion, although no line ratio
can support this choice, which is cross-checked after the DEM runs.
\citet{Brooks_etal:99} noticed that the 357.946/359.375\,\AA\
intensity ratio does not agree with theory, which they interpreted as
due to the blending of the \ion{Ne}{v}\,(357.946\,\AA) line with
\ion{Ne}{iv}. Also they outlined that the position patterns are not a
perfect match to either \ion{Ne}{v} or \ion{Ne}{iv}, casting a slight
doubt on the identifications. The DEM reconstrunction underestimates
this blend ($I_{\rm th}/I_{\rm obs} \sim 0.6$), supporting the
possibility that a further line may contribute to the observed
intensity. In Paper I it was suggested that the blend of \ion{Ne}{v}
and \ion{Ne}{iv} in the SERTS-89 spectrum was unlikely both because
the unblended \ion{Ne}{iv} line is in good agreement with observation
and because our adopted laboratory wavelengths indicate a separation
above the resolution of the instrument.  \citet{Brooks_etal:99} found
that the position pattern of \ion{Ne}{iv}\,(358.476\,\AA) has large
error bars and is more typical of the temperature of \ion{Fe}{x}. We
find that the \ion{Fe}{x} contribution is 10\,\% and this blend is
reproduced just above the observational error bars.  An examination of
the wavelength shifts in individual segments reveals some
inconsistencies between \ion{Ne}{v}\,(357.946\,\AA) and
\ion{Ne}{v}\,(359.375\,\AA) (especially in segment 3(4) where the
former is -0.26\,\AA\ away from the laboratory wavelength) and between
\ion{Ne}{v}\,(358.476\,\AA) and \ion{Ne}{v}\,(359.375\,\AA)(expecially
in segment 4(7L) where the former is 0.26\,\AA\ away from the
laboratory wavelength). \ion{Ne}{v}\,(359.375\,\AA), on the other
hand, does not reveal shifts larger than 0.1\,\AA, which supports our
identification. Such a situation suggest the presence of unidentified
lines contributing to the blend with \ion{Ne}{v}\,(358.476\,\AA) and
\ion{Ne}{v}\,(357.946\,\AA) produced possibly from lower temperature
plasma, given the agreement found in the SERTS-89 active region
spectrum.  The 2s$^2$\,2p$^2$\,$^1$D - 2s\,2p$^3$\,$^1$P transition
(365.599\,\AA) is thought to form a blend with \ion{Fe}{x}, which
\citet{Brooks_etal:99} found consistent with the position pattern
which is a mixture of lines of different temperature. The DEM
reconstruction is in good agreement with observation.

The 2s$^2$ 2p$^2$ $^3$P - 2s 2p$^3$ $^3$D transitions in \ion{Si}{ix}
were also observed in the NIS-1 wavelength range. The 1-1 and 1-2
components ($\lambda^{\rm ad}$\,=\,344.954\,\AA\ and 345.120\,\AA)
form a blend at 345.104\,\AA\ and the 2-1, 2-2 and 2-3 ($\lambda^{\rm
ad}$\,=\,349.620\,\AA, 349.791\,\AA\ and 349.860\,\AA) another at
$\lambda^{\rm cor}$\,=\,349.856\,\AA. \citet{Brooks_etal:99} found
that these 3 components have the same position patterns which fall in
the correct temperature class, with appropriate relative
intensities. All observed \ion{Si}{ix} lines were found in good
agreement with observations in the SERTS-89 spectrum (Paper I). The
only unblended component (0-1 at 341.950\,\AA) was assigned code B[1]
by \citet{Brooks_etal:99} and is chosen here in the integral inversion
to sample the region around $\log\,T\,=\,6.05$. Nevertheless, the two
blends are underestimated by the DEM, which is strikingly in contrast
with the results of Paper I and \citet{Brooks_etal:99}. We note that
the composite ratios are dependent on density below $\log\,N_{\rm
e}\,=\,9$. In the mean spectrum they would indicate $N_{\rm e}\sim 2
\times 10^8$ cm$^{-3}$, which is not consistent with the electron
density inferred from \ion{Mg}{viii}. Therefore, in this case the
discrepancies could be due to uncertainties in the dependence of the
population densities on density or to density inhomogeneities which
are not accounted for in the method used. Since this problem is not
present at higher densities, i.e. in the SERTS-89 active region, the
former explanation is preferred. The region of formation overlaps both
with \ion{Mg}{x} and \ion{Fe}{xi}.

The 2s$^2$ 2p$^2$ $^1$D$_2$ - 2s 2p$^3$ $^1$D$_2$ transition in
\ion{Mg}{vii}\,(319.018\,\AA) was identified by \citet{Brooks_etal:99}
with a good position pattern typical of its temperature. This line may
be blended with \ion{Ni}{xv}, but the SERTS-89 analysis in Paper I
could not give a definite answer on the relative contributions because
of the high density sensitivity of the \ion{Mg}{vii} transition. Its
intensity in the CDS spectrum is underestimated by the DEM, but a
significant contribution from \ion{Ni}{xv} seems unlikely in a QS
spectrum. For \ion{Mg}{vii}, the 2s$^2$ 2p$^2$ $^3$P - 2s 2p$^3$ $^3$P
transitions also lie in the NIS-1 wavelength
range. \citet{Brooks_etal:99} identified the 0-1 component
(363.749\,\AA), the 1-0 (365.162\,\AA), 1-2 (365.221\,\AA) and 1-1
(365.234\,\AA) components blended together, and the blend of the 2-2
(367.658\,\AA) and 2-1 (367.671\,\AA) components, resulting in three
lines of code C[1], B[1], and A[1], respectively, with the same
position pattern. The analysis of Paper I overestimated these lines,
outlining possible inaccuracies in the atomic data. The present
analysis supports such conclusions as it overestimates the mean
intensities by a factor of 2. We note also that the composite line
ratios are density sensitive below $\log\,N_{\rm e}\,=\,8$, but both
observed ratios are at the Boltzmann limit or slightly above.

\ion{Al}{viii} 2s$^2$ 2p$^2$ $^3$P - 2s 2p$^3$ $^3$P transitions were
identified by \citet{Brooks_etal:99} in the NIS-1 spectrum as a blend
of the 1-0 (325.280\,\AA), 1-2 (325.296\,\AA) and 1-1 (325.343\,\AA)
components, and a blend of the 2-2 (328.183\,\AA) and 2-1
(328.230\,\AA) components. The 0-1 component was not observed, as
expected since its intensity is below the sensitivity of the
instrument. We are unable to produce a reliable Gaussian fit of the
1-0, 1-2, 1-1 blend in the spectra analysed while the 2-2, 2-1 blend
is underestimated by the DEM analysis. Note, however, the large
uncertainties in the \element{Al} abundance (see
Sect.\,\ref{sec:abundance}).

\subsubsection{N-like sequence}
\label{sec:N-like}

The \ion{O}{ii} 2s$^2$ 2p$^3$ $^2$D - 2s 2p$^4$ $^2$P multiplet was
identified with the 3/2-1/2 component at 537.832\,\AA\ blended with
\ion{C}{iii} and the 5/2-3/2 (538.262\,\AA) and the 3/2-3/2
(538.320\,\AA) components blended together and with \ion{C}{iii}. As
discussed in Sect.\,\ref{sec:Be-like}, such blends may be affected by
line opacities and their formation temperature is below the trusted
temperature range for the DEM reconstruction. For these reasons and
because of a lack of suitable atomic data, the 2s$^2$ 2p$^3$ $^4$S -
2s$^2$ 2p$^2$ 3s $^4$P lines at 539.086-539.854\,\AA\ are not
discussed here. The 2s$^2$ 2p$^3$ $^2$D - 2s$^2$ 2p$^2$ 3s $^2$P lines
are identified as a blend of the 5/2-3/2 (616.303\,\AA) and the
3/2-3/2 (616.379\,\AA) components and a contribution of the 3/2-1/2
(617.063\,\AA) component to a \ion{O}{iv} blend. The former is
underestimated by the DEM and the latter gives a negligible
contribution to the \ion{O}{iv} blend.

The \ion{Ne}{iv} 2s$^2$\,2p$^3$\,$^2$D - 2s\,2p$^4$\,$^2$P and
2s$^2$\,2p$^3$\,$^4$S - 2s\,2p$^4$\,$^4$P transitions are
observed. The 2s$^2$\,2p$^3$\,$^2$D - 2s\,2p$^4$\,$^2$P lines (NIS-1)
are blended with other ions, the 3/2-1/2 (357.825\,\AA) with
\ion{Ne}{v} (see Sect.\,\ref{sec:C-like}) and the 5/2-3/2
(358.688\,\AA)and 3/2-3/2 (358.746\,\AA) with \ion{Fe}{xi}, and are
therefore excluded from the line ratio analysis. Both are
underestimated by the DEM, which indicates the presence of further
blends. \citet{Young_etal:98} suggested a contribution from
\ion{Si}{xi} and \ion{Fe}{xiv} to the blend with the 5/2-3/2 and
3/2-3/2 components. Since the contribution of \ion{Si}{xi} is found to
be negligible, and \ion{Fe}{xiv} is not expected to give a significant
contribution to the observed intensity in a QS spectrum, this
suggestion cannot be confirmed. The 2s$^2$\,2p$^3$\,$^4$S -
2s\,2p$^4$\,$^4$P transitions (541.126\,\AA, 542.070\,\AA,
543.886\,\AA), observed in NIS-2, are densitive insensitive and form
ratios in agreement with theory. The \ion{Ne}{iv}\,(543.886\,\AA) line
was used in the \citet{Brooks_etal:99} wavelength calibration and it
is selected here for the integral inversion. Good agreement is found
in the DEM analysis as well.
This situation could support the possibility that the discrepancies
found for the 2s$^2$\,2p$^3$\,$^2$D - 2s\,2p$^4$\,$^2$P lines may be
due to inaccuracies in the NIS-1/NIS-2 cross calibration as discussed
in Sect.\,\ref{sec:C-like} for the C-like \ion{O}{iii}
2s$^2$\,2p$^2$\,$^3$P - 2s$^2$\,2p\,3s\,$^3$P multiplet. This latter
possibility is contemplated because, as for the C-like case, the
region of formation is constrained by a \ion{Ne}{iv} NIS-2 line. As
for the C-like case, this is ruled out (see
Sect.\,\ref{sec:cross-calibration}).

The 2s$^2$\,2p$^3$\,$^2$D - 2s\,2p$^4$\,$^2$D multiplet in
\ion{Mg}{vi} (349.108-349.179\,\AA) lies unresolved in the NIS-1
wavelength range and is reported by \citet{Brooks_etal:99} as blended
with a \ion{Fe}{xi} line. However, the resulting position pattern did
not resemble either of the two ions, which resulted in assigning code
B[2] to this blend. The analysis of Paper I overestimated this blend
and it was suggested that, because of fitting problems, only two of
the four \ion{Mg}{vi} multiplet components were actually measured. The
present DEM analysis of the CDS spectrum overestimates this multiplet
as well and, given the lower resolution of the CDS instrument, the
explanation that only two components were actually fitted is no longer
sustainable. Given also the discrepancies in the position pattern, the
likely explanation seems now that the atomic data for \ion{Mg}{vi} are
in error.  Nevertheless, we are forced to use this multiplet to
constrain the \element{Mg}/\element{Ne} abundance ratio (see
Sect.\,\ref{sec:abundance}), obtaining consistent results.

The 2s$^2$ 2p$^3$ $^4$S - 2s 2p$^4$ $^4$P multiplet in \ion{Al}{vii}
lies in the NIS-1 wavelength range and the identification of its
components was discussed by \citet{Brooks_etal:99}.  The strongest
3/2-5/2 component (356.892\,\AA) was identified by
\citet{Brooks_etal:99} with a position pattern that could indicate a
weak unknown blend and led to the assignment of code C[2]. The DEM
reconstruction overestimates this line by 40\,\%.  The 3/2-3/2
component (353.777\,\AA) is thought to be blended with
\ion{Fe}{xiv}. The position pattern for this blend is more like that
of the \ion{Fe}{xiv} line formation temperature. Scaling from the
3/2-5/2 component of the \ion{Al}{vii} multiplet,
\citet{Brooks_etal:99} estimated, however, that the \ion{Fe}{xiv}
contribution should be about a third of the total intensity. We are
unable to produce a reliable Gaussian fit of this multiplet, but the
DEM reconstruction predicts a \ion{Fe}{xiv} contribution of about 1/5.
The weakest 1/2-1/2 component (352.159\,\AA) is expected to contribute
a few percent of a blend with \ion{Fe}{xii} and \ion{Mg}{v}. This is
confirmed by the DEM reconstruction and good agreement is found for
this blend (see Sects.\,\ref{sec:O-like} and \ref{sec:N-like}).

The 2s$^2$\,2p$^3$\,$^4$S - 2s\,2p$^4$\,$^4$P transitions in
\ion{Si}{viii} lie in the NIS-1 wavelength range. All three lines are
observed in the expected flux ratio. The 3/2-3/2 and 3/2-5/2 lines
(316.218\,\AA, 319.839\,\AA) have code A[1] and were used in the
\citet{Brooks_etal:99} wavelength calibration.  The DEM reconstruction
reproduces these lines, including the 3/2-1/2 component at
314.356\,\AA, within the observational uncertainties.  The region of
formation overlaps with \ion{Mg}{viii} which is used for the integral
inversion.

\subsubsection{O-like sequence}
\label{sec:O-like}

The \ion{Mg}{v} 2s$^2$ 2p$^2$ $^3$P - 2s 2sp$^5$ $^3$P multipet lies
in the NIS-1 wavelength range. The 2-2 component (353.092\,\AA) is
expected to be the strongest.  \citet{Thomas_Neupert:94} identified
five of the six lines, the 1-0 component not being reported. The 1-1
line was indicated as blended with \ion{Na}{vii}. The analysis of the
SERTS-89 data carried out in Paper\,I overestimated the 1-0 and 2-2
components, and reproduced the other four within the observational
uncertainites.  \citet{Brooks_etal:99} indicated that the 1-0 line
(352.197\,\AA) is dominated primarily by an \ion{Fe}{xii} line (used
in the wavelength calibration) as reflected in the position pattern
and intensity, and indicated a further blend with \ion{Al}{vii}. They
also indicated that the 0-1 line (354.221\,\AA) is more intense than
expected and could be blended. They ruled out a possible blend of the
1-1 component with \ion{Na}{vii}. For the unblended lines the position
patterns show large error bars so that these identifications were
indicated as not secure. In the analysis carried out here, only the
blend at $\lambda^{\rm obs}$\,=\,352.089\,\AA\ is considered since the
other \ion{Mg}{v} lines are too weak to be fitted reliably. Good
agreement is found, confirming the blend with \ion{Fe}{xii} and
\ion{Al}{vii}. The expected intensities for the other transitions of
this multiplet are reported in Tab.\,5.

\subsubsection{F-like sequence}

\citet{Brooks_etal:99} identified the 2s$^2$ 2p$^5$ $^2$P - 2s 2p$^6$
$^2$S doublet in \ion{Mg}{iv}. The 1/2-1/2 component (323.309\,\AA) is
too weak to be fitted reliably in our spectra and it is predicted to
have an intensity of 0.8 erg cm$^{-2}$ s$^{-1}$ sr$^{-1}$ in the mean
spectrum. The 3/2-1/2 component (320.995\,\AA) is blended with
\ion{Fe}{xiii} and their intensities are reproduced within the
observational uncertainties.

\subsubsection{Na-like sequence}
\label{sec:Na-like}

\citet{Brooks_etal:99} assigned code B[1] to both lines of the
3s\,$^2$S - 3p\,$^2$P transition in \ion{Ca}{x} (557.759\,\AA,
574.007\,\AA). Their ratio in the average spectrum is 1.5, which may
indicate a weak blend affecting the 574.007\,\AA\ line since $R_{\rm
th}\,=\,1.9$. The stronger line was used by \citet{Brooks_etal:99} in
the wavelength calibration and it is used here for the integral
inversion.  The DEM reconstruction underestimates the 574.007\,\AA\
line, except in segments 4(7L), 5(7U) and 7(8U) where it is reproduced
within the observational uncertainties, supporting the case of
blending affecting this line.

Regarding the 3p $^2$P - 3d $^2$D multiplet in \ion{Ar}{viii}, we fit
the 1/2-3/2 (519.464\,\AA) and 3/2-5/2 (526.496\,\AA) components as in
\citet{Brooks_etal:99}.  The 3/2-3/2 component (526.893\,\AA) is
predicted to have negligible intensity, consistent with the
observation. The 1/2-3/2 (519.464\,\AA) predicted intensity is only
12\,\% of the observed intensity, and therefore its identification is
doubtful.  The DEM reconstruction for the 3/2-5/2 (526.496\,\AA)
component is 50\,\% lower than observed.

\subsubsection{Mg-like sequence}

For this sequence, we consider the identification of
\ion{S}{v}\,(518.299\,\AA) (3s 3p $^1$P$_{1}$ - 3s 4s $^1$S$_{0}$) and
\ion{Ar}{vii}\,(585.754\,\AA) (3s$^2$ $^1$S$_{0}$ - 3s 3p
$^1$P$_{1}$).  For the former we obtain $I_{\rm th}/I_{\rm obs} \sim
0.15$.  Note that the \element{S} abundance is taken from
\citet{Feldman_etal:92} since no reliable estimate can be made with
the data at hand, and that \ion{S}{v}\,(518.299\,\AA) lies close to
the lower limit of the wavelength range of the NIS-2 detector.
\ion{Ar}{vii}\,(585.754\,\AA) is reproduced within the observational
uncertainties.

\subsubsection{Al-like sequence}

\citet{Brooks_etal:99} identified the 1/2-3/2 and 3/2-5/2 components
of the 3s$^2$ 3p $^2$P - 3s 3p$^2$ $^2$D multiplet in \ion{Ca}{viii}
(582.845\,\AA\ and 596.935\,\AA).  We are unable to obtain a reliable
Gaussian fit to the 596.935\,\AA\ line, but good agreement is found
with the observed 1/2-3/2 (582.845\,\AA) component. The 3/2-3/2
component (597.851\,\AA) is found to contribute 15\,\% to a blend with
\ion{O}{iii}.

The 1/2-3/2 and 3/2-5/2 components of the same multiplet in
\ion{Fe}{xiv} were identified, the latter indicated as blended with
\ion{Al}{vii}. We were unable to obtain reliable fits to the observed
lines in the set analysed here, but we predict a negligible
contribution to the blend with \ion{Al}{vii}. The predicted intensity
for the strongest component (1/2-3/2 at 334.178\,\AA) is only 1.2 erg
cm$^{-2}$ s$^{-1}$ sr$^{-1}$. In Paper I severe problems were found in
the atomic data for \ion{Fe}{xiv}, but these cannot be addressed here
because of the weakness of these lines in our QS spectra.

\subsubsection{Si-like sequence}

Lines of the \ion{Fe}{xiii} 3s$^2$ 3p$^2$ $^3$P - 3s 3p$^3$ $^3$D
multiplet were identified by \citet{Brooks_etal:99}. The blend of the
2-2 and 2-1 components is predicted to be much weaker than observed,
and therefore we do not confirm their identification at $\lambda^{\rm
cor}$\,=\,372.131\,\AA. The 0-1 component, identified at $\lambda^{\rm
cor}$\,=\,348.148\,\AA, is overestimated by 50\,\% and therefore we
cannot confirm the \citet{Brooks_etal:99} suggestion that this line
has an unknown blend. The blend of the 1-2 and 1-1 components,
identified by \citet{Brooks_etal:99} at $\lambda^{\rm
cor}$\,=\,359.777\,\AA, is reproduced within the observational
uncertainties.

The \ion{Fe}{xiii} 3s$^2$ 3p$^2$ $^3$P - 3s 3p$^3$ $^3$P multiplet was
also considered by \citet{Brooks_etal:99}. The strongest component is
the 1-1 (312.095\,\AA) which we predict to have an intensity of 3.5
erg cm$^{-2}$ s$^{-1}$ sr$^{-1}$ in the mean spectrum considered here,
and was not detected by \citet{Brooks_etal:99}. The 2-1 component
(321.394\,\AA) was identified at $\lambda^{\rm cor}$\,=\,321.436\,\AA\
and is here predicted to have an intensity of 1.8 erg cm$^{-2}$
s$^{-1}$ sr$^{-1}$ in our mean spectrum although we were unable to
obtain a reliable fit to the observed line because it is too weak. The
2-2 component (320.800\,\AA) contributes 15\,\% of the intensity of
the blend with \ion{Mg}{iv}, which is reproduced within the
observational uncertainties. The 1-2 component (311.552\,\AA),
identified by \citet{Brooks_etal:99} at $\lambda^{\rm
cor}$\,=\,311.476\,\AA, is predicted to have negligible intensity in
our mean spectrum. Note that lines of \ion{Fe}{xiii} were
systematically underestimated in Paper I, which was ascribed to
uncertainties in the atomic modelling for \element{Fe}.

Lines of \ion{Ca}{vii} were also identified by \citet{Brooks_etal:99},
but we could not obtain reliable Gaussian fits and report the
predicted intensities only in Table\,5.

\subsubsection{P-like sequence}
\label{sec:P-like}

Lines of \ion{Fe}{xii} are observed in the NIS-1 wavelength range.
\citet{Brooks_etal:99} assigned code A[1] to the unblended
\ion{Fe}{xii}\,(364.468\,\AA) line (3s$^2$\,3p$^3$\,$^4$S$_{3/2}$ -
3s\,3p$^4$\,$^4$P$_{5/2}$). The DEM reconstruction of this line is
within the observational error bars.  The other components of the
multiplet, 3/2-1/2 at 346.852\,\AA\ and 3/2-3/2 at 352.107\,\AA, have
codes B[2] and B[1], respectively.  The 352.107\,\AA\ line was
reported as blended with \ion{Mg}{v} and \ion{Al}{vii}.  The predicted
intensities for the 3/2-1/2 (346.852\,\AA) line and for the blend of
the 3/2-3/2 (352.107\,\AA) component with \ion{Mg}{v} and
\ion{Al}{vii} are in good agreement with observations.  The
346.852/364.468\,\AA\ ratio, which is density insensitive, is in good
agreement with theory. The DEM reconstruction for these lines is also
in good agreement with observations.  \citet{Brooks_etal:99}
identified also the 3s$^2$ 3p$^3$ $^2$D$_{5/2}$ - 3s 3p$^4$
$^2$D$_{5/2}$ (338.263\,\AA) line. The DEM analysis, however, does not
confirm the identification since it predicts an intensity a factor 10
weaker than observed.  The region of formation overlaps with
\ion{Si}{x}.

\subsubsection{S-like sequence}
\label{sec:S-like}

All components of the 3s$^2$\,3p$^4$\,$^3$P - 3s\,3p$^5$\,$^3$P
multiplet in \ion{Fe}{xi} were identified in \citet{Brooks_etal:99}
and the 2-2 component (352.661\,\AA) was used in the wavelength
calibration (code A[2]).  The 1-0 component (349.046\,\AA) is blended
with \ion{Mg}{vi}, but, according to the DEM reconstruction, the
\ion{Fe}{xi} contribution to the intensity is negligible.  The 0-1
component (358.621\,\AA) is blended with \ion{Ne}{iv}, but further
blending is required to match the observed intensity. The \ion{Si}{xi}
line at 358.656\,\AA\ is found to have negliglible contribution to
this blend in QS conditions.  A possible blend with an unknown
component was suggested by \citet{Brooks_etal:99} for the 2-1
component at 341.113\,\AA. We find the 356.519/352.661\,\AA\ and
369.154/352.661\,\AA\ line ratios in agreement with theory, but the
341.113/352.661\,\AA\ line ratio is approximately a factor 2.7 too
high, which confirms the blending of the 341.113\,\AA\ line. The DEM
analysis supports such conclusions too. The contribution of the
\ion{Fe}{xiii} line at 356.59\,\AA\ is found to be negligible.  The
region of formation overlaps with \ion{Si}{ix} and \ion{Mg}{xi}.

\subsubsection{Cl-like sequence}
\label{sec:Cl-like}

The \ion{Fe}{x} 3s$^2$\,3p$^5$\,$^2$P$_{3/2}$ -
3s\,3p$^6$\,$^2$S$_{1/2}$ transition at 345.723\,\AA\ has code A[1] in
\citet{Brooks_etal:99} and was used in the wavelength calibration.
Good agreement is found with the DEM reconstruction.  The other
component of the multiplet, the 1/2-1/2 at 365.543\,\AA, is blended
with \ion{Ne}{v}. The DEM reconstruction for this blend is in good
agreement with observation.  The line at $\lambda^{\rm
cor}$\,=\,358.424\,\AA\ was considered to be a blend of \ion{Ne}{v}
and the 3s$^2$ 3p$^4$ 3d $^4$F$_{9/2}$ - 3s 3p$^5$ 3d $^4$F$_{9/2}$
\ion{Fe}{x} transition by \citet{Brooks_etal:99}. The DEM analysis
indicates that the \ion{Fe}{x} contribution to this blend is
10\,\%. The temperature of formation of \ion{Fe}{x} overlaps with
\ion{Mg}{ix}.  Note that \ion{Fe}{x} lines were systematically
underestimated in Paper I, which was attributed to an inaccurate
treatment of the ionisation/recombination balance.

%
\subsection{The mean DEM and the DEM from individual segments}
\label{sec:DEM_segments}
%

\begin{figure*}[ht]
\hbox to \hsize {\hss \epsfxsize=160mm \epsfbox {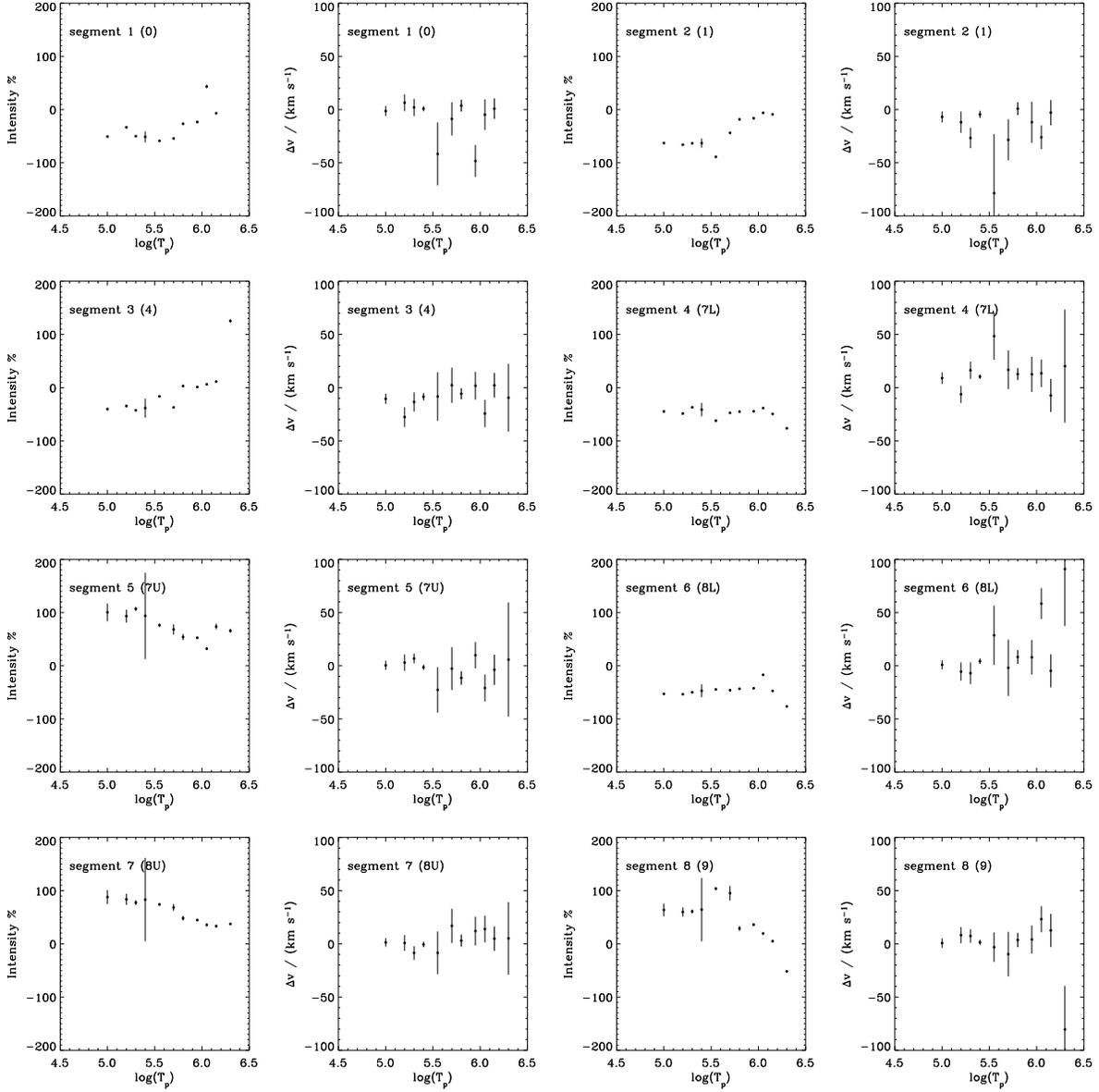}\hss}
   \caption{Intensity variation compared to the mean spectrum and
   wavelength shift in velocity units with respect to the mean
   spectrum of the lines selected for the integral inversion in each
   segment.  Error bars are evaluated combining the measurement errors
   as in Sect.\,\ref{sec:rad_cal}. Data points with no bars have
   uncertainties below the size of the symbols. Note that in segments
   1(0) and 2(1) we are unable to obtain a reliable Gaussian fit of
   the \ion{Si}{xii}\,(520.662\,\AA) and therefore this line is not
   reported for such segments. }
   \label{fig:cdsdem_intvel}
\end{figure*}

The intensities of the lines selected for the integral inversion in
each segment are compared with the mean intensity in
Fig.\,\ref{fig:cdsdem_intvel}. Also in Fig.\,\ref{fig:cdsdem_intvel}
we compare the line shift with respect to the average spectrum for
each segment. Line intensities in the transition region tend to be
lower than average when velocities of the order of 30 km s$^{-1}$,
either as upflows or downflows, are present. In segments 1(0), 2(1),
4(7L), and 6(8L) velocities of the order of 50 km s$^{-1}$ are
present. In such conditions the plasma can travel a hydrostatic
scale-height in 300 s at $\log\,T\,=\,5.5$. This is an order of
magnitude shorter than the relaxation time-scales for the ground
states of Ne, Mg, Si, and S at the inferred electron densities (see
Paper I) and therefore we expect ionisation equilibrium to break down
in such regions.  Note, however, that the relaxation time-scales for
ground levels presented in Paper I are evaluated directly from the
ionisation/recombination cross-sections. The time required to reach
ionisation equilibrium can be much shorter than these values if
departures are small.  Hydrodynamic calculations \citep[see,
e.g.,][]{Spadaro_etal:03, Lanza_etal:01} indicate that such departures
for a single loop may last only for short times and may be not too
large.


\begin{figure*}[ht]
\hbox to \hsize {\hss \epsfxsize=130mm \epsfbox {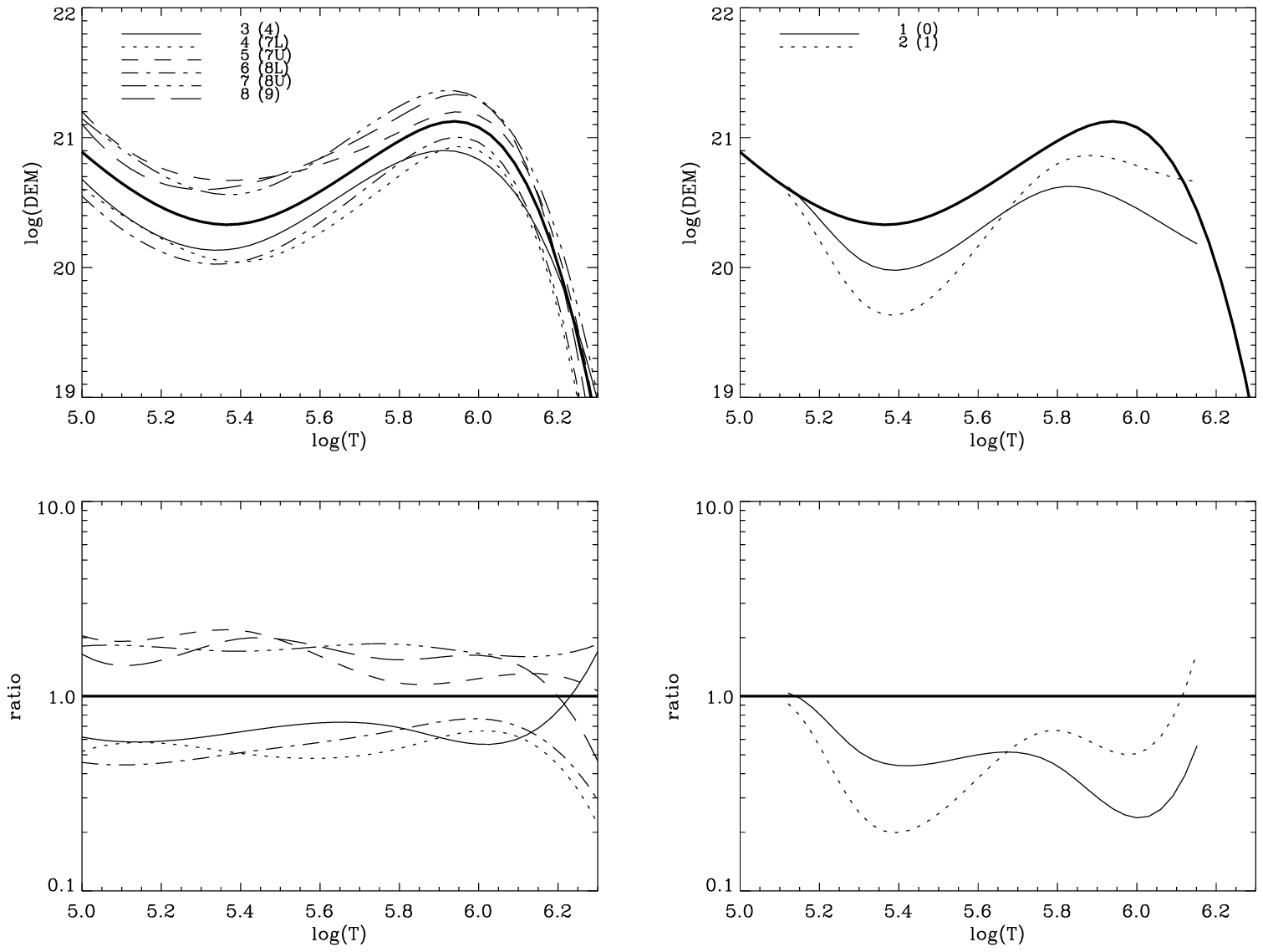}\hss}
   \caption{DEM from the selected segments compared with the mean
     spectrum (thick solid line). Above $\log\,T\,=\,6.0$ the DEM is
     less reliable because of the weakness of the lines. The ratio of
     each segment's DEM with the mean spectrum DEM (lower panels) show
     that the plasma distribution in temperature is essentially the
     same in all segments except 1(0) and 2(1). These latter show the
     largest downflow velocities (see Fig.\,\ref{fig:cdsdem_intvel}).}
   \label{fig:cdsdem_demplot}
\end{figure*}

The DEM from the mean spectrum and from the selected segments are
shown in Fig.\,\ref{fig:cdsdem_demplot}. Above $\log\,T\,=\,6.0$ the
DEM is less reliable because of the weakness of the lines in the quiet
Sun. The ratios of each segment's DEM with the mean spectrum DEM
(lower panels of Fig.\,\ref{fig:cdsdem_demplot}) show that the
distribution of the plasma in temperaure is essentially the same in
all segments except those around the extended brightening (1(0) and
2(1)). These latter are the regions with the lowest electron density
which also show {\it downflow} velocities of the order of 50 km
s$^{-1}$ (Fig.\,\ref{fig:cdsdem_intvel}) and are expected to contain
plasma not in ionisation equilibrium. It is not possible, therefore,
to establish whether the different DEM shape is real or due to a
breakdown of the assumption of ionisation equilibrium.
If real, such a shape may suggest that the two extended \ion{Mg}{ix}
brightening segments correspond to the top of systems of extended
loops whose footpoints are outside the selected segments. In this
case, transition region lines may form in compact loops which do not
reach coronal temperatures and are thermally decoupled from the plasma
emitting in \ion{Mg}{ix}. Such compact loops are known to show high
velocities \citep[see, e.g.,][]{Lanzafame_etal:99}.
Interestingly, the DEM for segment 4(7L), which correspond to a
\ion{Mg}{ix} dark region with $N_{\rm e} \approx 10^8$ cm$^{-3}$, does
not show evident deviations in the DEM shape, despite {\it upflow}
velocities of the order of 50 km s$^{-1}$ being present.

There are other possible indications of deviations from ionisation
equilibrium. Comparing the Li-like \ion{Mg}{x}\,(624.95\,\AA) with the
Be-like \ion{Mg}{ix}\,(368.057\,\AA) in each segment (as well as the
Li-like \ion{Al}{xi}\,(550.04\,\AA) with the Be-like
\ion{Al}{x}\,(332.788\,\AA), see Sects.\,\ref{sec:Li-like} and
\ref{sec:Be-like}) we have shown that the ratio $I_{\rm th}/I_{\rm
obs}$ for the former is systematically higher than for the latter.
Such a systematic behaviour can be interpreted as being due to an {\it
overionisation} of the Li-like stage with respect to the equilibrium
ionisation balance. It is expected that this behaviour appears evident
in our analysis because of the use of \element{Si} lines in the
calculation of the DEM at the formation temperatures of these lines
($\log\,T\,\sim 6.0$ and above), which have relaxation time-scales for
the ground states 50\,\% shorter than those of \element{Mg} at such
temperatures (see Paper I).  In this case, in fact, any effect which
is dependent on the ionisation/recombination rates of elements with
longer relaxation time-scales than \element{Si} would become
detectable.  The \ion{Mg}{x}\,(624.95\,\AA) /
\ion{Mg}{ix}\,(368.057\,\AA) line intensity ratio is not correlated
with line shifts and this could be considered an evidence against
non-equilibrium conditions. It can be expected, however, that small
scale events could cause some degree of non-equilibrium in the corona
without producing bulk flow of plasma and therefore measurable line
shifts.  As discussed in Sects.\,\ref{sec:Li-like} and
\ref{sec:Be-like}, an alternative explanation is that recombination to
the 2s 2p $^1$P excited term of the Be-like ions or ionisation to it
from e.g. the 2s 2p$^2$ $^4$P term in the B-like ions give a
contribution which is not accounted for in our analysis since
level-resolved ionisation balance calculations for \element{Mg} and
\element{Al} are not yet available.

%
\subsection{Elemental abundances}
\label{sec:abundance}
%

\begin{figure}[ht]
\hbox to \hsize {\hss \epsfxsize=88mm \epsfbox {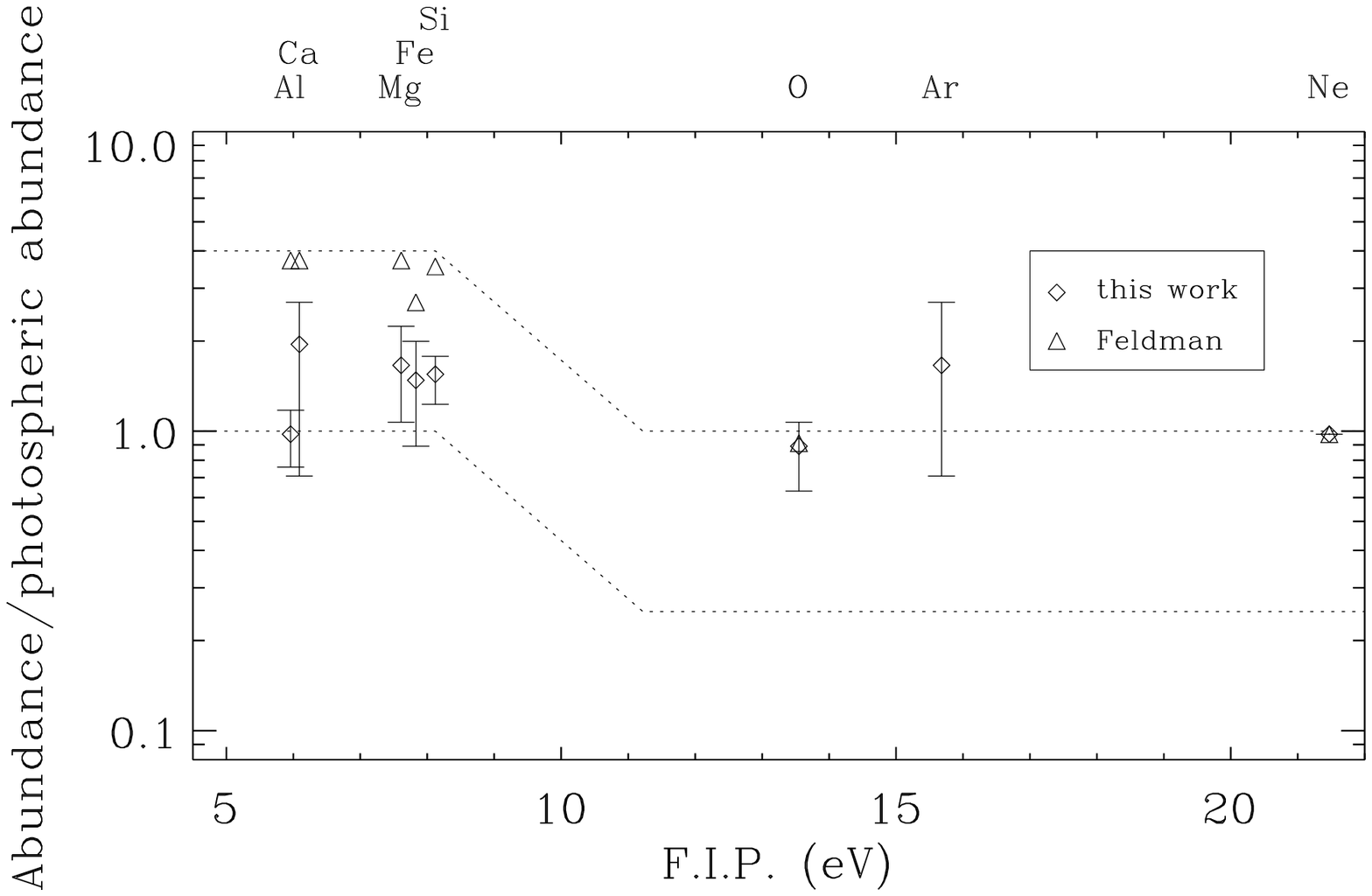}\hss}
    \caption{ Derived elemental abundances for our CDS s6011r00
	    observation compared with the \citet{Feldman_etal:92}
	    coronal abundances. Abundances are plotted as ratios with
	    the photospheric values taken from \citet{Feldman:92}.
            The 3-$\sigma$ confidence interval is also reported,
	    obtained for all elements except \element{Mg} by a Monte
	    Carlo simulation. The 3-$\sigma$ confidence interval for
	    \element{Mg} is obtained by propagation of errors via the
	    \element{Mg}/\element{Ne} constraint (see text). 
	    \element{Ne} is taken as reference.}
   \label{fig:abundance}
\end{figure}

In evaluating elemental abundances, we start by assuming that the Ne
abundance (with respect to H) in the transition region and corona is
the same as in the photosphere. In other words, we assume that
high-FIP elements (i.e. elements with their first ionisation potential
- FIP - greater than approximately 10 eV) have coronal abundances
close to their photospheric values. This assumption is necessary since
we are unable to obtain elemental abundances relative to H using the
CDS-NIS spectrum alone, and therefore our abundances must be taken
relative to one of the elements emitting in the CDS-NIS
spectrum. Alternatively, we could have assumed, without loss of
validity of our results, that low-FIP elements have photospheric
abundances and that high-FIP elements are depleted in the corona. In
this latter case, all abundances would have been scaled by a constant
factor (approximately from the higher dotted line to the lower in
Fig.\,\ref{fig:abundance}) and the DEM would have been multiplied by
the same factor.
 
We have selected all observed lines emitted by ions other than Ne
which do not show evident blending or problems in the atomic data and
then evaluated the elemental abundances by a $\chi^2$ minimization for
a given DEM. The DEM has therefore been re-evaluated with the new
abundances and the procedure iterated until the maximum correction to
the abundances was below 10\,\%.  This procedure allows a good
determination of the abundance ratios of elements whose lines form at
close temperatures. In such cases, the uncertainties in the DEM
reconstruction tend to cancel out and the dominant uncertainties on
the abundance ratios derive from the atomic data and observational
uncertainties.

In the DEM analysis carried out here, however, the low temperature
range ($5.0 \le \log\,T_{\rm e}\,\le\,5.8$) is sampled by high FIP
elements and the high temperature range ($5.8 < \log\,T_{\rm
e}\,\le\,6.0$) by low FIP elements. As a consequence, above
$\log\,T_{\rm e}\,=\, 5.8$ variations of abundances of the low FIP
elements cannot be effectively disentangled from variations of the DEM
with temperature, although the relative abundances amongst low FIP
elements remains well constrained.

\addtocounter{table}{+1}

\begin{table*}[ht]
\caption[]{Abundance enhancement factors (with respect to photospheric
           composition) for the low FIP elements in the selected
           segments. $\beta$ is the FIP bias deduced from the observed
           \ion{Mg}{vi}\,(349.2\,\AA)/\ion{Ne}{vi}\,(562.8\,\AA) line
           intensity ratio as in \citet{Young_Mason:97}. The result
           for \element{Ar} in segment 2(1) is not reported because
           the estimated error is very large.}
\label{tab:fip_segments}
\begin{center}
\begin{tabular}{llcccccccc}
\hline
Region               & Segment & \multicolumn{7}{c}{Enhancement factor} & $\beta$ \\
                     &         & \element{Al}        & \element{Ca}        & \element{Mg}        & \element{Fe}        & \element{Si}        & \element{O}         & \element{Ar}     &  \\
\hline
Extended Brightening & 1(0)    & 0.5$_{-0.2}^{+0.2}$ & 1.9$_{-1.5}^{+0.9}$ & 1.7$_{-0.3}^{+0.3}$ & 1.2$_{-0.5}^{+0.6}$ & 1.5$_{-0.3}^{+0.2}$ & 0.6$_{-0.2}^{+0.2}$ & 2.1$_{-0.9}^{+0.9}$ & 3.6 \\
                     & 2(1)    & 1.1$_{-0.5}^{+0.6}$ & 1.7$_{-1.0}^{+1.9}$ & 1.9$_{-0.3}^{+0.3}$ & 2.2$_{-0.7}^{+0.7}$ & 2.3$_{-0.5}^{+0.5}$ & 0.6$_{-0.1}^{+0.2}$ &                     & 3.5 \\
\\
Narrow brightening   & 3(4)    & 1.1$_{-0.4}^{+0.4}$ & 1.7$_{-1.1}^{+1.5}$ & 1.7$_{-0.3}^{+0.3}$ & 2.0$_{-0.6}^{+0.8}$ & 1.7$_{-0.5}^{+0.4}$ & 0.9$_{-0.1}^{+0.2}$ & 1.8$_{-0.8}^{+1.2}$ & 3.5 \\
                     & 5(7U)   & 1.0$_{-0.4}^{+0.5}$ & 3.7$_{-1.1}^{+1.9}$ & 1.5$_{-0.3}^{+0.3}$ & 0.9$_{-0.3}^{+0.3}$ & 1.0$_{-0.2}^{+0.2}$ & 0.9$_{-0.1}^{+0.2}$ & 1.5$_{-0.7}^{+1.0}$ & 3.2 \\
                     & 7(8U)   & 0.8$_{-0.5}^{+0.4}$ & 4.8$_{-1.8}^{+2.8}$ & 1.7$_{-0.3}^{+0.3}$ & 1.4$_{-0.3}^{+0.5}$ & 1.6$_{-0.3}^{+0.4}$ & 1.0$_{-0.2}^{+0.2}$ & 0.5$_{-0.3}^{+0.6}$ & 3.2 \\
                     & 8(9)    & 0.9$_{-0.6}^{+0.5}$ & 2.1$_{-0.6}^{+0.9}$ & 1.7$_{-0.3}^{+0.3}$ & 0.7$_{-0.2}^{+0.3}$ & 1.2$_{-0.3}^{+0.2}$ & 1.0$_{-0.2}^{+0.2}$ & 1.3$_{-0.6}^{+0.8}$ & 3.1 \\
\\
Dark regions         & 4(7L)   & 1.1$_{-0.8}^{+0.8}$ & 2.6$_{-1.0}^{+1.4}$ & 0.8$_{-0.3}^{+0.3}$ & 0.9$_{-0.4}^{+0.4}$ & 1.3$_{-0.3}^{+0.3}$ & 0.7$_{-0.1}^{+0.2}$ & 2.3$_{-1.3}^{+0.8}$ & 1.5 \\
                     & 6(8L)   & 0.6$_{-0.5}^{+0.4}$ & 2.3$_{-1.7}^{+1.4}$ & 1.2$_{-0.3}^{+0.3}$ & 0.6$_{-0.2}^{+0.2}$ & 1.1$_{-0.2}^{+0.2}$ & 0.9$_{-0.2}^{+0.2}$ & 2.1$_{-1.7}^{+1.2}$ & 2.6 \\
\hline
\end{tabular}
\end{center}
\end{table*}

In order to overcome such difficulties, we have used an additional
constraint based on the
\ion{Mg}{vi}\,(349.2\,\AA)/\ion{Ne}{vi}\,(562.8\,\AA) line intensity
ratio. The peaks of temperature of formation for \ion{Mg}{vi} and
\ion{Ne}{vi} are essentially the same, and this attractive
characteristic has been widely used in deriving coronal
\element{Mg}/\element{Ne} abundances \citep[e.g.,][ and references
therein]{Widing_Feldman:01}, often with \element{Ne} used as a proxy
for the high FIP group of elements and \element{Mg} for the low FIP
group of elements. The
\ion{Mg}{vi}\,(349.2\,\AA)/\ion{Ne}{vi}\,(562.8\,\AA) line intensity
ratio observed by CDS-NIS has been used, for instance, by
\citet{Young_Mason:97} to derive \element{Mg}/\element{Ne} abundance
variations in a recently emerged flux region.

Note that neither \ion{Mg}{vi}\,(349.2\,\AA) nor
\ion{Ne}{vi}\,(562.8\,\AA) are reproduced satisfactorily by the DEM
analysis (see Sects.\,\ref{sec:N-like} and \ref{sec:B-like}), the most
likely cause being inaccuracies in the atomic modelling. Lacking other
trustworthy constaints, however, we impose the criterion that the
solution reproduces the observed
\ion{Mg}{vi}\,(349.2\,\AA)/\ion{Ne}{vi}\,(562.8\,\AA) line intensity
ratio within 10\,\%.

The results for the mean spectrum are shown in
Fig.\,\ref{fig:abundance} together with the \citet{Feldman_etal:92}
coronal abundances for comparison. In Fig.\,\ref{fig:abundance},
abundances are shown as ratios to their photosperic values
\citep[taken from][]{Feldman:92}.  The 3-$\sigma$ confidence interval
is also reported, obtained for all elements except \element{Mg} by a
Monte Carlo simulation. The 3-$\sigma$ confidence interval for
\element{Mg} is obtained by propagation of errors via the
\element{Mg}/\element{Ne} constraint discussed above. We remind the
reader that \element{Ne} is taken as reference.

In the mean spectrum all elements are found close to photospheric
composition, with enhancement factors below two except for
\element{Ca} whose abundance is 2.1 photospheric. The results for
\element{Al} and \element{Ca}, however, must be considered with
caution. In fact, using lines of \ion{Al}{x} and \ion{Ca}{x} only, for
instance, the derived abundance for \element{Al} and \element{Ca}
would be about four times photospheric, and therefore close to coronal
values. Including all lines listed in Table\,\ref{dem_line_list} we
derive abundances close to photospheric for \element{Al} and
\element{Ca} too. A likely explanation for this situation is that the
accuracy of the atomic data for some of the ions of \element{Al} and
\element{Ca} are still below the requirements for this type of
analysis. An alternative explanation is that there is some enrichment
process taking place, which affects some of the ionisation stages of
the elements with lowest FIP more than others. The derived abundance
for the high FIP element \element{Ar} is consistent with photospheric
composition; the large error bar is due to the fact that only one
\element{Ar} line can be used for the analysis. The abundance of
\element{O} is everywhere very close to photospheric composition
(except in segment 1(0) for which our analysis is uncertain) with a
3-$\sigma$ uncertainty of typically 20\,\%.

The same procedure has been applied to the spectra from the selected
segments (Table\,\ref{tab:fip_segments}). We find that the abundance
of all elements remains essentially photospheric, except that of
\element{Ca}, as discussed above for the mean spectrum. As in the mean
spectrum, the abundance of \element{Ar} has large error bars,
especially in segment 2(1).

Following \citet{Widing_Feldman:01} we indicate with $\beta$ the FIP
bias based on the \element{Mg}/\element{Ne} abundance ratio relative
to the photospheric compositon (i.e. $\beta\,=\,1$ indicates
photospheric composition). We derive $\beta$ from the observed
\ion{Mg}{vi}\,(349.2\,\AA)/\ion{Ne}{vi}\,(562.8\,\AA) line intensity
ratio as in \citet{Young_Mason:97}. Table\,\ref{tab:fip_segments}
indicates that $\beta\,\approx\,3$ in all segments (as well as in the
mean spectrum) except in the dark \ion{Mg}{ix} regions, where $\beta$
is closer to unity. This comparison, on the one hand, confirms that
the \element{Mg} abundance is essentially constant from segment to
segment; on the other hand, it shows that the
\element{Mg}/\element{Ne} abundance ratio inferred from the observed
\ion{Mg}{vi}\,(349.2\,\AA)/\ion{Ne}{vi}\,(562.8\,\AA) line intensity
ratio differs from the DEM results by a factor of 2. This is due to
the underlying assumption of a constant DEM in the line ratio
technique combined with the temperature (and density) dependence of
the \ion{Mg}{vi}\,(349.2\,\AA)/\ion{Ne}{vi}\,(562.8\,\AA) line
intensity ratio.  An assessment of the accuracy of the atomic
population modelling for \ion{Mg}{vi} and \ion{Ne}{vi} is, however,
required to make a sound comparison between the two techniques.

\subsection{NIS-1/NIS-2 cross-calibration uncertainties}
\label{sec:cross-calibration}

\begin{table}[h]
\caption[]{Comparison of the median $I_{\rm th}/I_{\rm obs}$ 
           between NIS-1
           and NIS-2 lines for the ions \ion{O}{iii}, \ion{Ne}{iv}, 
           and \ion{Ne}{v}}
\label{table:NIS-1-2_comparison}
\begin{center}
\begin{tabular}{lrr}
\hline
              & \multicolumn{2}{c}{MEDIAN $I_{\rm th}/I_{\rm obs}$} \\
Ion           & NIS-1 & NIS-2 \\
\hline
\ion{O}{iii}  & 0.2   & 0.9   \\
\ion{Ne}{iv}  & 0.6   & 1.2   \\
\ion{Ne}{v}   & 0.9   & 1.6   \\
\hline
\end{tabular}
\end{center}
\end{table}

\citet{Landi_etal:97}, considering ions whose lines are observed in
both NIS channels, found discrepancies that they attributed to errors
in the NIS-1/NIS-2 cross-calibration.  They derived a relative
calibration from in-flight data and this together with the preliminary
results from the laboratory calibration provided a basis for Version 1
of the CDS NIS intensity calibration. As detailed in
\citet{Lang_etal:02}, the calibration has been revised since then and
the burn-in corrections, particularly with the adoption of the
wide-slit burn-in correction, are now improved. Thus without repeating
their work with the present day data reduction and calibration
software it is very difficult to make any comparisons. However, the
revisions to the calibration since their work have substantially
reduced the discrepancies.  In our analysis, lines of \ion{O}{iii},
\ion{Ne}{iv}, and \ion{Ne}{v} show a mismatch between NIS-1 and NIS-2
too.  In fact, for such ions, it looks like either NIS-1 lines are
underestimated or NIS-2 lines are overestimated (or both) by the DEM
reconstruction (see Sects.\,\ref{sec:B-like}, \ref{sec:C-like}, and
\ref{sec:N-like}).  Similarly to the \citet{Landi_etal:97} analysis,
the correction factor implied by such lines is rather large, being
between a factor 2 and 4, and greater than the uncertainties assigned
to the intensity calibration\footnote{Version 4 of the CDS calibration
is used in the present analysis, along with the data reduction as
detailed in \citet{Lang_etal:00}.}.

\begin{figure}[ht]
\hbox to \hsize {\hss \epsfxsize=88mm \epsfbox {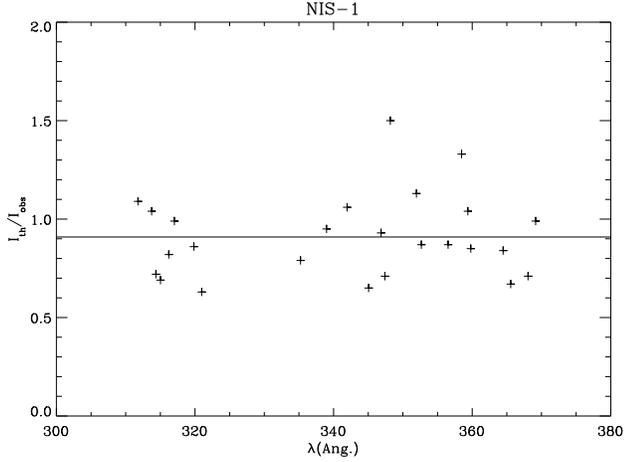}\hss}
   \caption{$I_{\rm th}/I_{\rm obs}$ ratio for NIS-1 lines with 
            $(\sigma_0 \chi_i)^2 \le 3$, excluding also lines 
            affected by unknown blends or large
            uncertainties in the atomic data (crosses), 
	    together with a $I_{\rm th}/I_{\rm obs}\,=\,{\rm const.}$ 
	    least-square fit.}
   \label{fig:qfr_N1}
\end{figure}

We have therefore performed some numerical experiments in which the
responsitivity of one of the two channels has been multipled by an
arbitrary correction factor to take into account possible errors in
the NIS-1/NIS-2 cross-calibration.  We found that a correction of the
NIS-1/NIS-2 relative calibration greater than approximately 20\,\%
increases the reconstruction errors for the lines used in the integral
inversion (labelled ``i'' in Table\,\ref{dem_line_list} and
5) and for those chosen for the comparison in the
forward sense (labelled ``f'' in Table\,\ref{dem_line_list} and
5) well above the observational uncertainties.
Such a correction would therefore produce inconsistencies amongst
those lines which have been selected in both channels because of their
reliability.  Note that, although the $G$-functions of these lines
peak at different temperatures\footnote{To improve the conditioning of
the problem, the use of lines with similar $G$-functions in the
integral inversion is avoided.}, such inconsistencies show up because
the overlapping part of the $G$-functions still constrains the DEM
substantially.  In the {\em data adapive smoothing approach} used here
(see Paper I for a description) the automatic smoothing tends to
adjust the $\lambda$-parameter in response to the change in the
relative calibration to fit both NIS-1 and NIS-2 lines in the list
used for the integral inversion, still maintaining the smoothness of
the DEM.  In the present case, where lines from both channels are
mixed in the list, if we introduce an increasing correction factor in
the NIS-1/NIS-2 relative calibration, we have first a distortion of
the DEM with the theoretical intensities increasingly failing to
reproduce the observations, then either the automatic smoothing fails
or the solution becomes unacceptable. Using an arbitrary smoothing
method, it would still be possible to reproduce the intensities with
an oscillatory DEM with multiple peaks, like that computed by
\citet{Landi_Landini:97}, but such a solution is not justified by the
information content of the data.

\begin{figure}[h]
\hbox to \hsize {\hss \epsfxsize=88mm \epsfbox {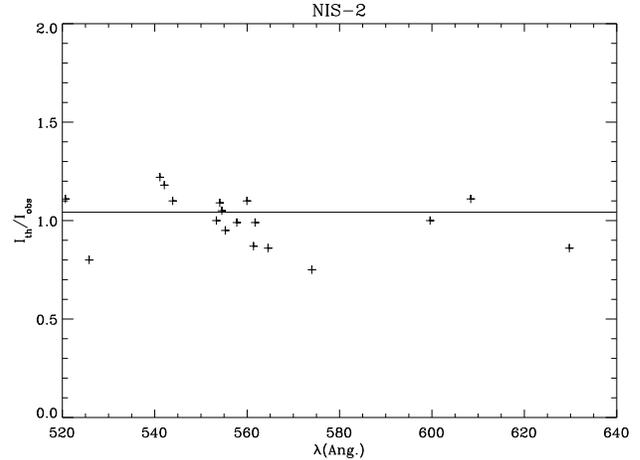}\hss}
   \caption{Same as Fig.\,\ref{fig:qfr_N1} for NIS-2 lines.  From the
            comparison with the previous plot we infer that the error
            in the NIS-1/NIS-2 cross calibration is at most 10\,\%.}
   \label{fig:qfr_N2}
\end{figure}

In Figs.\,\ref{fig:qfr_N1} and \ref{fig:qfr_N2} we plot the $I_{\rm
th}/I_{\rm obs}$ ratio for lines with $(\sigma_0 \chi_i)^2 \le 3$
(excluding also lines affected by unknown blends or large
uncertainties in the atomic data), together with a $I_{\rm th}/I_{\rm
obs} \,=\, {\rm const.}$ least-square fit on both NIS-1 and NIS-2
data. Comparing the two fits, we infer that the uncertainty in the
cross-calibration is at most 10\,\%. This should be considered only an
upper limit, given the high rejection value adopted for $(\sigma_0
\chi_i)^2$.

We therefore conclude that the discrepancies seen in \ion{O}{iii},
\ion{Ne}{iv}, and \ion{Ne}{v} cannot be ascribed to errors in the
cross-calibration of the NIS channels.

\section{Conclusions}
\label{sec:conclusions}

This work builds on the spectral analysis of \citet{Brooks_etal:99},
exploiting the DEM in the spectral analysis of a quiet Sun area
observed by SOHO-CDS and addressing the issue of how the DEM of small
subareas (indicated here as segments of the whole raster sequence)
combine to give the DEM of a larger area. The DEM analysis is also
exploited to improve the line identification performed by
\citet{Brooks_etal:99} and to investigate possible elemental abundance
variations from region to region.

The distribution of the plasma in temperaure is found to be
essentially the same in all segments except those around the extended
\ion{Mg}{ix} brightening (1(0) and 2(1)). These latter are the regions
with the lowest electron density and {\it downflow} velocities of the
order of 50 km s$^{-1}$ (Fig.\,\ref{fig:cdsdem_intvel}), which are
expected to contain plasma not in ionisation equilibrium. This
situation is interpreted here as indicating that, in regions where
$\log\,N_{\rm e}\,>\,7.3$ and downflow velocities are less than 50 km
s$^{-1}$, radiative cooling and (or) conductive terms dominate(s) the
energy balance in the QS upper transition region and inner corona.  In
such conditions the DEM differs only by a roughly constant factor
depending on the total amount of heating in that region.  The segments
around the extended \ion{Mg}{ix} brightening might correspond to a
combination of the tops of large loops whose footpoints lie outside
the selected segments and of compact loops which do not reach coronal
temperature and are thermally decoupled from the plasma emitting in
\ion{Mg}{ix}. Such compact loops are known to show high velocities
\citep[see, e.g.,][]{Lanzafame_etal:99}.  The analysis carried out in
this work shows that complex structure in the DEM due to
inhomogeneities, such as multiple peaks, is not justified by the data
and is not expected unless uncertainties affect large areas of the
region analysed.

The elemental abundances are found to be close to photospheric
composition and do not show significant variations from region to
region. Only the \element{Ca} abundance seems to be enhanced with
respect to photospheric values with some variation from region to
region, but the results are quite uncertain since the scatter of
spectral lines of the \element{Ca} ions is rather large. The
\element{Al} abundance is also rather uncertain for the same
reasons. Note that the use of \ion{Al}{x} and \ion{Ca}{x} only would
lead to estimated abundances about four times photospheric in the mean
spectrum, and a photospheric composition is inferred only using all
lines listed in Table\,\ref{dem_line_list}.

Two indications of departures from ionisation equilibrium are
found. The first is that when $\log\,N_{\rm e}\,<\,7.3$ and the velocity
exceeds 50 km s$^{-1}$ the DEM reconstruction has higher
uncertainties. In such conditions the plasma can travel a hydrostatic
scale-height in 300 s at $\log\,T\,=\,5.5$, which is an order of
magnitude shorter than the relaxation time-scales for the ground
states at the inferred electron densities.  This affects the
calculation of the DEM over an area of approximately 2 by 80
arcsecs$^2$, but the effect is less important on the calculation of
the DEM based on the spectrum averaged over 20 by 240 arcsecs$^2$.
The second indication comes from the comparison between the Li-like
\ion{Mg}{x}\,(624.95\,\AA) and the Be-like
\ion{Mg}{ix}\,(368.057\,\AA) in each segments, which points to an
overabundance of the Li-like stage (longer relaxation time-scale for
the ground state) with respect to equilibrium conditions.  This is
supported also by the comparison between the Li-like
\ion{Al}{xi}\,(550.04\,\AA) and the Be-like
\ion{Al}{x}\,(332.788\,\AA), which show the same systematic behaviour
as the \ion{Mg}{x} and \ion{Mg}{ix} lines.  A viable alternative
explanation of this latter behaviour is, however, that recombination
to the excited term 2s 2p $^1$P of the Be-like ions or ionisation to
it from e.g. the 2s 2p$^2$ $^4$P term in the B-like ions give a
contribution which is not accounted for in our analysis since
level-resolved ionisation balance calculations for \element{Mg} and
\element{Al} are not yet available.

We note that the relaxation time-scales for ground levels presented in
Paper I are evaluated directly from the ionisation/recombination
cross-sections. The time required to reach the ionisation equilibrium
can be much shorter than these values if departures are
small. Hydrodynamic calculations \citep[see, e.g.,][]{Spadaro_etal:03,
Lanza_etal:01} indicate that such departures for a single loop may
last only for short times and may be not too large. The analysis
presented here indicates that departures from ionisation equilibrium
may significantly affect the analysis on small areas (of the order of
2 by 80 arcsec$^2$ in the present observations), but when considering
an average on larger areas (of the order of 20 by 240 arcsec$^2$ in
the present observations) they tend to be of less importance because
most of the plasma appears to be close to ionisation
equilibrium. Departures from ionisation equilibrium, therefore, may
not invalidate the DEM analysis on large areas, allowing us to derive
a distribution in temperature for the bulk plasma in
quasi-equilibrium. For such areas the DEM still represents a useful
support for the spectral analysis, but accurate results could not
always be obtained because of the presence of relatively small
transients of short duration.

Significant discrepancies are generally found for density sensitive
lines. Some multiplet that were reproduced within the observational
uncertainties in the SERTS-89 active region analysis (Paper I) are not
reproduced accurately in the present analysis because they have larger
density sensitivity in the range of electron densities involved
here. This is likely to be caused by the assumption of uniform
electron density (or uniform electron pressure) and a bivariate DEM
(differential in temperature and density) approach would be required.

Accurate results are also prevented by the uncertanties in the atomic
modelling of some ions, which confuses the picture further.  Because
of this, we are sometimes unable to find with confidence the cause of
the observed discrepancies. Unfortunately, such uncertainities would
prevent also a reliable bivariate DEM approach, and therefore the
possibility of taking into account unresolved inhomogeneities of the
observed plasma.  In some instances, e.g. \ion{O}{iv}, \ion{Ne}{vi},
\ion{Mg}{vi}, \ion{Si}{x}, and \ion{Mg}{vii}, our analysis reveals
unambiguously errors in the atomic data.

\begin{acknowledgements}
CDS was built and is operated by a consortium led by the Rutherford
Appleton Laboratory and which includes the Mullard Space Science
Laboratory, the NASA Goddard Space Flight Center, Oslo University and
the Max-Planck-Institute for Extraterrestrial Physics, Garching. SOHO
is a mission of international cooperation between ESA and NASA.  This
work was done in the context of the ADAS project.  The authors thank
an anonymous referee for useful comments on the original manuscript.
\end{acknowledgements}

\bibliographystyle{apj}

\begin{thebibliography}{43}
\expandafter\ifx\csname natexlab\endcsname\relax\def\natexlab#1{#1}\fi

\bibitem[{Brooks {et~al.}(1999)Brooks, Fischbacher, Fludra, et al.}]{Brooks_etal:99}
Brooks, D.~H., Fischbacher, G.~A., Fludra, A., et al. 1999, A\&A, 347,
  277

\bibitem[{Dere {et~al.}(1997)Dere, Landi, Mason, Monsignori~Fossi, \&
  Young}]{Dere_etal:97}
Dere, K.~P., Landi, E., Mason, H.~E., Monsignori~Fossi, B.~C., \& Young, P.~R.
  1997, A\&AS, 125, 149

\bibitem[{Edl\'{e}n(1983a)}]{Edlen:83a}
Edl\'{e}n, B. 1983a, Phys. Scripta, 28, 51

\bibitem[{Edl\'{e}n(1983b)}]{Edlen:83b}
---. 1983b, Phys. Scripta, 28, 48

\bibitem[{Edl\'{e}n(1984)}]{Edlen:84}
---. 1984, Phys. Scripta, 30, 135

\bibitem[{Edl\'{e}n(1985a)}]{Edlen:85a}
---. 1985a, Phys. Scripta, 31, 345

\bibitem[{Edl\'{e}n(1985b)}]{Edlen:85b}
---. 1985b, Phys. Scripta, 32, 59

\bibitem[{Edl\'{e}n(1985c)}]{Edlen:85c}
---. 1985c, Phys. Scripta, 32, 86

\bibitem[{Fawcett(1975)}]{Fawcett:75}
Fawcett, B.~C. 1975, Atomic Data and Nuclear Data Tables, 16, 138

\bibitem[{Feldman(1992)}]{Feldman:92}
Feldman, U. 1992, Physica Scripta, 46, 202

\bibitem[{Feldman {et~al.}(1992)Feldman, Mandelbaum, Seely, Doschek, \&
  Gursky}]{Feldman_etal:92}
Feldman, U., Mandelbaum, P., Seely, J.~F., Doschek, G.~A., \& Gursky, H. 1992,
  ApJS, 81, 387

\bibitem[{Harrison(1997)}]{Harrison:97}
Harrison, R.~A. 1997, Sol. Phys., 175, 467

\bibitem[{Harrison {et~al.}(2003)Harrison, Harra, Brkovic, \&
  Parnell}]{Harrison_etal:03}
Harrison, R.~A., Harra, L.~K., Brkovic, A., \& Parnell, C.~E. 2003, A\&A, 755

\bibitem[{Harrison {et~al.}(1999)Harrison, Lang, Brooks, \&
  Innes}]{Harrison_etal:99}
Harrison, R.~A., Lang, J., Brooks, D.~H., \& Innes, D.~E. 1999, A\&A, 351, 1115

\bibitem[{Harrison {et~al.}(1995)Harrison, Sawyer, Carter, et al.}]{Harrison_etal:95}
Harrison, R.~A., Sawyer, E.~C., Carter, M.~K., et al. 1995, Sol. Phys.,
  162, 233

\bibitem[{Innes {et~al.}(1997)Innes, Inhester, Axford, \&
  Wilhelm}]{Innes_etal:97}
Innes, D.~E., Inhester, B., Axford, W.~I., \& Wilhelm, K. 1997, Nature, 386,
  811

\bibitem[{Jup\'en {et~al.}(1993)Jup\'en, Isler, \& Tr\"abert}]{Jupen_etal:93}
Jup\'en, C., Isler, R.~C., \& Tr\"abert, E. 1993, MNRAS, 264, 627

\bibitem[{Kelly(1987)}]{Kelly:87}
Kelly, R.~L. 1987, J. Phys. Chem. Ref. Data 16, Supp. 1

\bibitem[{Landi \& Bhatia(2003)}]{Landi_Bhatia:03}
Landi, E. \& Bhatia, A.~K. 2003, ApJ, 589, 1075

\bibitem[{Landi \& Landini(1997)}]{Landi_Landini:97}
Landi, E. \& Landini, M. 1997, A\&A, 327, 1230

\bibitem[{Landi {et~al.}(1997)Landi, Landini, Pike, \& Mason}]{Landi_etal:97}
Landi, E., Landini, M., Pike, C.~D., \& Mason, H.~E. 1997, Sol. Phys., 175, 553

\bibitem[{Landi {et~al.}(2004)Landi, Storey, \& Zeippen}]{Landi_etal:04}
Landi, E., Storey, P.~J., \& Zeippen, C.~J. 2004, ApJ, 607, 640

\bibitem[{Lang {et~al.}(2001)Lang, Brooks, O'Mullane, et al.}]{Lang_etal:01}
Lang, J., Brooks, D.~H., O'Mullane, M.~G., et al. 2001, Sol. Phys., 201,
  37

\bibitem[{Lang {et~al.}(2000)Lang, Kent, Breeveld, et al.}]{Lang_etal:00}
Lang, J., Kent, B.~J., Breeveld, A.~A., et al. 2000, J. Opt. A, 2, 88

\bibitem[{Lang {et~al.}(2002)Lang, Thompson, Pike, Kent, \&
  Foley}]{Lang_etal:02}
Lang, J., Thompson, W.~T., Pike, C.~D., Kent, B.~J., \& Foley, C.~R. 2002, in
  The Radiometric Calibration of SOHO. ISSI Scientific Report SR-002, ed.
  A.~Pauluhn, M.~C.~E. Huber, \& R.~von Steiger, 105

\bibitem[{Lanza {et~al.}(2001)Lanza, Spadaro, Lanzafame, et al.}]{Lanza_etal:01}
Lanza, A.~F., Spadaro, D., Lanzafame, A.~C., et al. 2001, ApJ, 547, 1116

\bibitem[{Lanzafame {et~al.}(2002)Lanzafame, Brooks, Lang, Summers, Thomas, \&
  Thompson}]{Lanzafame_etal:02}
Lanzafame, A.~C., Brooks, D.~H., Lang, J., Summers, H.~P., Thomas, R.~J., \&
  Thompson, A.~M. 2002, A\&A, 384, 242

\bibitem[{Lanzafame {et~al.}(1999)Lanzafame, Spadaro, Consoli, Marsch, \&
  Brooks}]{Lanzafame_etal:99}
Lanzafame, A.~C., Spadaro, D., Consoli, L., Marsch, E., \& Brooks, D.~H. 1999,
  in The 8th SOHO workshop: Plasma Dynamics and Diagnostics in the Solar
  Transition Region and Corona, ed. J.-C. Vial \& Kaldeich-Sch\"urman, 429

\bibitem[{Martin {et~al.}(1995)Martin, Sugar, Musgrove, \&
  Dalton}]{Martin_etal:95}
Martin, W.~C., Sugar, J., Musgrove, A., \& Dalton, G.~R. 1995, NIST Database
  for Atomic Spectroscopy, Version 1.0, NIST Standard Reference Data base 61

\bibitem[{McWhirter \& Summers(1984)}]{McWhirter_Summers:84}
McWhirter, R. W.~P. \& Summers, H.~P. 1984, in Applied Atomic Collision Physics
  Vol. 2, ed. C.~F. Barnett \& M.~F.~A. Harrison, Vol.~2 (Academic Press), 51

\bibitem[{Moore(1993)}]{Moore:93}
Moore, C.~E. 1993, in Tables of Spectra of Hydrogen, Carbon, Nitrogen and
  Oxygen Atoms and Ions, ed. J.~W. Gallacher, CRC Series in Evaluated Data in
  Atomic Physics (CRC Press)

\bibitem[{Shirai {et~al.}(1990)Shirai, Funataka, Mori, Sugar, Wiese, \&
  Nakai}]{Shirai_etal:90}
Shirai, T., Funataka, Y., Mori, K., Sugar, J., Wiese, W.~L., \& Nakai, Y. 1990,
  J. Chem. Phys. Ref. Data, 19, 127

\bibitem[{Spadaro {et~al.}(2003)Spadaro, Lanza, Lanzafame, et al.}]{Spadaro_etal:03}
Spadaro, D., Lanza, A.~F., Lanzafame, A.~C., et al. 2003, ApJ, 582, 486

\bibitem[{Summers(1994)}]{Summers:94}
Summers, H.~P. 1994, ADAS manual, JET Joint Undertaking, JET-IR(94)07

\bibitem[{Summers(2001)}]{Summers:01}
---. 2001, The ADAS manual, version 2-3, http://adas.phys.strath.ac.uk

\bibitem[{Thomas \& Neupert(1994)}]{Thomas_Neupert:94}
Thomas, R.~J. \& Neupert, W.~M. 1994, ApJS, 91, 461

\bibitem[{Thompson(1990)}]{Thompson:90}
Thompson, A.~M. 1990, A\&A, 240, 209

\bibitem[{Thompson(1991)}]{Thompson:91}
---. 1991, in Intensity Integral Inversion Techniques: a Study in Preparation
  for the SOHO Mission, ed. R.~A. Harrison \& A.~M. Thompson, RAL Report No.
  RAL91-092

\bibitem[{Widing \& Feldman(2001)}]{Widing_Feldman:01}
Widing, K.~G. \& Feldman, U. 2001, ApJ, 555, 426

\bibitem[{Wiese(1985)}]{ORNL:85}
Wiese, W.~L., ed. 1985, Spectroscopic Data for Iron, ORNL No. 6089/V4 (Oak
  Ridge National Laboratory)

\bibitem[{Young {et~al.}(2003)Young, Del~Zanna, Landi, Dere, Mason, \&
  Landini}]{Young_etal:03}
Young, P.~R., Del~Zanna, G., Landi, E., Dere, K.~P., Mason, H.~E., \& Landini,
  M. 2003, ApJS, 144, 135

\bibitem[{Young {et~al.}(1998)Young, Landi, \& Thomas}]{Young_etal:98}
Young, P.~R., Landi, E., \& Thomas, R.~J. 1998, A\&A, 329, 291

\bibitem[{Young \& Mason(1997)}]{Young_Mason:97}
Young, P.~R. \& Mason, H.~E. 1997, Sol. Phys., 175, 523

\end{thebibliography}

\end{document}